\documentclass[
    reprint,
    amsmath,
    amssymb,
    aip,
    jcp
]{revtex4-2}

\usepackage{graphicx}
\usepackage{dcolumn}
\usepackage{bm}
\usepackage{xcolor}
\usepackage{ragged2e}
\usepackage{braket}
\usepackage{placeins}
\usepackage[T1]{fontenc}

\newcommand{\Tr}{\mathrm{Tr}}
\newcommand{\id}{\mathbb{I}}
\newcommand{\dd}{\mathrm{d}}

\begin{document}


\title{Floquet Dressing and Bath Spectral Effects on the Geometric Phase of a Driven Dissipative Qubit}

\author{Chirag Arora}
\email{chirag13@bu.edu}
\email{chemchirag13@gmail.com}

\affiliation{Department of Chemistry, Boston University, Boston, Massachusetts 02215, USA}

\date{\today}


\begin{abstract}

The geometric phase (GP) of a periodically driven symmetric
spin-boson model (SBM) is investigated using the process-tensor
time-evolving matrix product operator (PT-TEMPO) method. A
longitudinal monochromatic field drives the qubit along the \emph{same}
coordinate that couples to an Ohmic bosonic environment. The
gauge-invariant kinematic GP is evaluated from the reduced-state
trajectory, and a circular non-additivity measure quantifies the
combined effects of driving and dissipation relative to
driven-unitary and undriven-dissipative references. Amplitude,
frequency, coupling-strength, and bath-cutoff sweeps reveal regimes
in which periodic driving reduces or enhances environmental
phase deformation relative to the unitary counterpart. The one-cycle GP exhibits pronounced nonmonotonic drive frequency dependence, while its bath-induced shift
depends sensitively on both coupling strength and cutoff.
These trends are interpreted qualitatively in terms of Floquet
dressing and bath spectral structure. A two-cycle echo provides
a complementary reversibility diagnostic, whose persistence
signal does not generally measure the Tong GP. The results
characterize the non-additive influence of driving and dissipation
on geometric-phase dynamics and identify parameter regimes of
reduced bath-induced deformation.

\end{abstract}



\maketitle


\section{Introduction}
\label{sec:introduction}

Geometric phases (GPs) \cite{cohen2019geometric,pancharatnam1956generalized,longuet1958studies} arise when a quantum system undergoes an evolution whose accumulated phase depends on the geometry of the path traversed by the quantum state rather than solely on the duration of the evolution. Since Berry's original formulation for adiabatic cyclic evolution of isolated quantum systems,\cite{berry1984quantal} GPs have become a central concept in quantum mechanics\cite{zwiebach2022mastering,chruscinski2012geometric} with applications in condensed matter physics,\cite{xiao2010berry,nandy2018berry,mikitik1999manifestation,shapere1989geometric,yan2025effect} chemical dynamics,\cite{resta2000manifestations,bouakline2018unambiguous,de1996dynamical,kendrick1997geometric,mead1992geometric,ryabinkin2017geometric,makri2023electronic} and quantum information science.\cite{yale2016optical,leek2007observation,magazzu2018asymptotic,murta2020berry} For a closed quantum system undergoing cyclic adiabatic evolution, the GP is determined by the closed trajectory traced by the state in its Hilbert space. More general formulations extend the concept to nonadiabatic and noncyclic evolution.\cite{aharonov1987phase,mukunda1993quantum1,mukunda1993quantum2,sjoqvist2000geometric,lombardo2006geometric,lombardo2015correction} In open quantum systems (OQSs), however, environmental interactions modify both the purity of the quantum state and deforms the trajectory followed by the system relative to its unitary counterpart. Consequently, the dissipative GP is sensitive to environment-induced deformations of the trajectory of OQSs.\cite{breuer2002theory,weiss2012quantum} The influence of an environment on GP has attracted considerable attention \cite{tong2004kinematic,wang2026dissipative,garrison1988complex,sinitsyn2009stochastic,mccaul2018driving,makri1997universal,viotti2023geometric} because GPs are often proposed as robust quantities and may be useful for emerging quantum technologies such as quantum batteries,\cite{cristiano2025geometric,hadipour2026spatial,kheiri2026dynamical} quantum computation,\cite{song2017continuous,wikeckowski2020majorana,leibfried2003experimental,vedral2003geometric} and as well as fundamental aspects of quantum synchronization,\cite{daniel2023geometric,wachtler2024topological,sun2026geometric} and cavity quantum electrodynamics (cQED).\cite{zheng2004unconventional,liu2017superadiabatic,lin2009robust}

GP effects are not merely a theoretical abstraction but have been experimentally observed in nonlinear frequency conversion,\cite{karnieli2022geometric} interferometric optical measurements,\cite{samlan2016study,yan2026probing} ultrafast experiments,\cite{yong2019valley,valahu2023direct} and higher-harmonic spectroscopies.\cite{yao2025abnormal,luu2018measurement,bai2024probing} In addition to these experimental signatures, several proposed experiments based on ultrafast techniques have also shown promise for tracking GP signatures in real time.\cite{otterpohl2024coherent,keefer2020visualizing,ye2025unraveling,mi2023geometric}

Periodically driven dissipative quantum systems \cite{hanggi1998driven,restrepo2016driven,goldman2014periodically,taft1998effects,mccaul2018driving,gamba2025open} have been studied in various contexts such as dynamical stabilization,\cite{makarov1995control,makri2001localization,makri1997stabilization} optimizing emission from double quantum wells,\cite{dong2004optimizing} quantum stochastic resonance (QSR),\cite{dong2004quantum,grifoni1996nonlinear,chen2023nonmarkovian} strong-field effects,\cite{meier1999non} drive--bath interference,\cite{zhao2010interference,cangemi2019dissipative} quantum control of entanglement,\cite{creffield2007quantum} and Floquet effects.\cite{mickiewicz2026benchmarking,mickiewicz2026exact,engelhardt2020dynamical,mori2023floquet,sato2025floquet} The effects of driving on the dissipative GP remain relatively unexplored, with only a few studies addressing this problem.\cite{villar2020geometric,fujikawa2009geometric,sun2009feedback,carollo2004spin,chu1989density,obada2014geometric,yang2026geometric}

In a recent study, Wang \textit{et al.}~\cite{wang2026dissipative} showed that the dissipative GP of an undriven SBM can be suppressed by \emph{static} inversion-symmetry breaking, reflecting a competition between the dynamical evolution of the qubit and environment-induced localization. Here, a periodically driven symmetric SBM is studied as a minimal model for examining the effect of \emph{dynamical} periodic inversion-symmetry breaking in an otherwise symmetric SBM. In particular, the drive and dissipative environment are intentionally chosen to couple to the \emph{same} system coordinate. The resulting dynamics depend not only on the bath spectral density, but also on the drive-induced quasienergy structure and the Fourier-resolved matrix elements of the system--bath coupling operator. This provides a direct setting in which to examine the connection between drive-induced Floquet restructuring, bath spectral sampling, and the resulting deformation of the GP.

The central question of this work is: \textbf{\emph{How does Floquet restructuring of the system, change the spectral channels through which the bath deforms the GP?}}

To explicitly quantify the \emph{non-additive} effect of driving and dissipation on the GP, a difference measure $\delta\gamma_D(t)$ is introduced (\emph{vide infra}). Using the inclusion-exclusion principle, this quantity measures the extent to which the drive dressed system's trajectory is distorted by dissipation beyond the separate contributions of the undriven dissipative and driven unitary dynamics respectively. The results are interpreted using complementary theoretical concepts such as Floquet theory \cite{magazzu2018asymptotic,mori2023floquet,grifoni1999dissipation} which describes the quasienergy structure generated by the periodic drive, coherent destruction of tunneling (CDT) \cite{grossmann1991coherent,kayanuma2008coherent,gong2009many,hai2013transparent,lu2011coherent,makri1997stabilization} which provides a mechanism for drive-induced suppression of effective tunneling at select $A/\Omega$ ratios and ideas from quantum stochastic resonance (QSR) which provide insight dynamical cooperation of drive and dissipation.\cite{makri2001localization,makarov1995control,dong2004optimizing,dong2004quantum,grifoni1996coherent,hanze2021quantum,chen2023nonmarkovian,grifoni1996nonlinear} 

The reduced density matrix (RDM) of the driven qubit is calculated using the numerically exact PT-TEMPO method.\cite{fux2024oqupy,fux2022process} For bilinearly coupled harmonic system--bath models,\cite{weiss2012quantum,breuer2002theory} this approach expresses the Feynman--Vernon influence functional \cite{feynman2000theory} as a tensor network.\cite{guo2024efficient,lacroix2026tensor} This allows the reduced dynamics to be propagated over timescales exceeding the bath correlation time while retaining nonlocal memory effects. PT-TEMPO is one of several tensor-network extensions \cite{bose2022multisite,kilda2020tensor,link2026tensor,link2024open,bose2023quantumdynamics} of the computationally demanding but powerful family of quasi-adiabatic path-integral (QuAPI) methods pioneered by Makri and coworkers.\cite{makri1995numerical,makri2023quantum} Following the prescription of kinematic GP introduced by Tong \textit{et al.}~\cite{tong2004kinematic}, the instantaneous eigenvalues and eigenvectors of the RDM are used to obtain the dissipative GP. This provides access to GP accumulation under non-unitary quantum evolution without assuming weak system--bath coupling, Markovian dynamics, or a particular perturbative master-equation ansatz.\cite{breuer2002theory,weiss2012quantum}

Following the general logic of geometric-phase interferometry in OQSs, \cite{viotti2023geometric} an additional test of the operational accessibility of the dissipative geometric phase, a two-cycle interferometric echo protocol is considered to assess environmentally modified reversibility and the limitations of relating echo persistence to the independently calculated kinematic geometric phase. 

The results demonstrate that bath-induced GP deformation depends jointly on the drive-induced Floquet transition frequencies, the corresponding Fourier-resolved system--bath matrix elements, and the environmental spectral weight available at those frequencies. This mechanism, referred to below as \emph{Floquet spectral steering}, provides a unified interpretation of the dependence of the GP on drive amplitude, drive frequency, bath cutoff frequency, and system--bath coupling strength.The remainder of this article is organized as follows. Section~\ref{sec:theory} introduces the key theoretical concepts.
Section~\ref{sec:floquet_picture} develops the Floquet spectral interpretation. Section~\ref{sec:results} presents the numerical results.
Summary and outlook is presented in Sec.~\ref{sec:summary}.


\section{THEORY}
\label{sec:theory}


\subsection{Driven spin-boson model (DSBM)}
\label{sec:DSBM}

A symmetric two-level system is described by the Hamiltonian
\begin{equation}
H_S^{(0)} = -\frac{\Delta}{2}\sigma_x
\label{eq:bare_system_hamiltonian}
\end{equation}
where $\Delta$ defines the intrinsic tunneling energy scale. In the
$\sigma_z$ basis, the two diabatic states are energetically degenerate,
while the tunneling term mixes these states and generates coherent
population oscillations on the characteristic system timescale $\tau_S \sim \Delta^{-1}$.
Because the Hamiltonian contains no static $\sigma_z$ bias, the two
localized diabatic states are energetically equivalent, and the system
possesses inversion symmetry under
$\sigma_z\rightarrow-\sigma_z$. The superscript "0" denotes the \emph{bare} (bath free) system in the absence of the drive field. Throughout, $\hbar=k_B=1$ (numerically), and energies, frequencies, and inverse times are expressed in units of \(\Delta\) unless otherwise specified 

A classical monochromatic field is applied longitudinally,
\begin{equation}
H_D(t)
=
-\frac{A}{2}\cos(\Omega t)\sigma_z
\label{eq:drive_hamiltonian}
\end{equation}
where $A$ and $\Omega$ are the drive amplitude and frequency,
respectively. The driven system Hamiltonian is therefore
\begin{equation}
H_S(t)
=
-\frac{\Delta}{2}\sigma_x
-
\frac{A}{2}\cos(\Omega t)\sigma_z
\label{eq:driven_system_hamiltonian}
\end{equation}

The instantaneous longitudinal field breaks the inversion symmetry of
the bare qubit. Because its sign reverses after half a drive period,
however, the periodically driven Hamiltonian retains the generalized
dynamical symmetry under $\mathcal{P}=\sigma_x$ since,
\begin{equation}
\mathcal{P}
H_S\left(t+\frac{T_D}{2}\right)
\mathcal{P}^{-1}
=
H_S(t)
\qquad
T_D=\frac{2\pi}{\Omega}
\label{eq:dynamical_inversion_symmetry}
\end{equation}

The environment is represented by a harmonic bath,
\begin{equation}
H_B
=
\sum_k\omega_k b_k^\dagger b_k
\label{eq:bath_hamiltonian}
\end{equation}
coupled to the qubit through a bilinear interaction
\begin{equation}
H_{SB}
=
-\frac{1}{2}\sigma_z B_z
\qquad
B_z
=
\sum_k c_k(b_k^\dagger+b_k)
\label{eq:system_bath_interaction}
\end{equation}

The complete driven spin-boson (DSBM) Hamiltonian is consequently
\begin{equation}
H(t)
=
-\frac{\Delta}{2}\sigma_x
-\frac{A}{2}\cos(\Omega t)\sigma_z
-\frac{1}{2}\sigma_z B_z
+\sum_k\omega_k b_k^\dagger b_k
\label{eq:total_hamiltonian}
\end{equation}

The bath is characterized by the exponentially truncated Ohmic
spectral density
\begin{equation}
J(\omega)
=
2\alpha\omega e^{-\omega/\omega_c}
\label{eq:spectral_density}
\end{equation}
where $\alpha$ controls the system--bath coupling strength and
$\omega_c$ sets the characteristic bath frequency scale.

The dynamics therefore involve three characteristic frequency scales,
$\Delta$, $\Omega$, and $\omega_c$, associated with the intrinsic
system dynamics, external drive, and dissipative environment,
respectively. The instantaneous effective field associated with the driven
Hamiltonian is
\begin{equation}
\mathcal{B}(t)
=
\left(
\Delta,
0,
A \cos(\Omega t)
\right)
\qquad
H_S(t)
=
-\frac{1}{2}\mathcal{B}(t)\cdot\boldsymbol{\sigma}
\label{eq:effective_field}
\end{equation}
so that the drive periodically changes both the amplitude and the direction of the effective field experienced by the qubit. Although the drive has zero time-averaged bias, it modifies the instantaneous Hamiltonian and consequently induces micromotion that affects the trajectory of the Bloch vector. This time-dependent
modification of the state trajectory provides the basis for influencing the bath induced deformation of the GP.


\subsection{PT-TEMPO parameters}
\label{sec:tempo}

The reduced density matrix (RDM) of the qubit is obtained from the dynamics generated by
Eq.~\eqref{eq:total_hamiltonian} by tracing over the bath degrees
of freedom,\cite{leggett1987dynamics,weiss2012quantum,breuer2002theory}
\begin{equation}
\rho(t)
=
\Tr_B
\left[
\rho_{\mathrm{tot}}(t)
\right]
\label{eq:reduced_density_matrix}
\end{equation}
where $\rho_{\mathrm{tot}}(t)$ denotes the density matrix of the combined driven qubit-bath system and is in general non-separable at $ t > 0 $.

The initial state of the total system is taken to be factorized as
\cite{weiss2012quantum,breuer2002theory}
\begin{equation}
\rho_{\mathrm{tot}}(0)
=
\rho(0)\otimes\rho_B(0;T)
\label{eq:factorized_initial_state}
\end{equation}
where $\rho(0)$ is the initial reduced state of the qubit and
$\rho_B(0;T)$ is the thermal equilibrium state of the bath at
temperature $T$. Explicitly, the thermal bath density operator is given by
\begin{equation}
\begin{aligned}
\rho_B(0;T)
=
\frac{e^{-\beta H_B}}
{\operatorname{Tr}_{B}\!\left[e^{-\beta H_B}\right]}
=
\frac{e^{-\beta H_B}}{Z_B},
\qquad
\beta=\frac{1}{k_B T}
\label{eq:thermal_bath_state}
\end{aligned}
\end{equation}
where
\begin{equation}
Z_B
=
\operatorname{Tr}_{B}
\left[
\exp(-\beta H_B)
\right]
\end{equation}
is the canonical partition function of the bath.

Unless otherwise stated, the PT-TEMPO calculations are performed
using $\alpha=0.05$ and $T/\Delta=1.5$, with representative bath
cutoff frequencies $\omega_c/\Delta\in\{1,5, 10, 20\}$. The drive parameters are varied as specified in Sec.~\ref{sec:results}. 
\\
The production calculations use time step of $dt=0.0025\,\Delta^{-1}$, memory time of $t_{\mathrm{mem}}=7\omega_c^{-1}$ and compression threshold of $\epsilon_{\mathrm{rel}}=10^{-6}$. Numerical convergence with respect to $dt$, $t_{\mathrm{mem}}$, and $\epsilon_{\mathrm{rel}}$ is discussed in Appendix~\ref{app:convergence}.


\subsection{Dissipative geometric phase (GP) and drive bath non-additivity}
\label{sec:gp}

It is convenient to parametrize the RDM of the qubit in the Bloch form as
\begin{equation}
\rho(t)
=
\frac{1}{2}
\left[
\mathbb{I}
+
\mathbf{r}(t)\cdot\bm{\sigma}
\right]
\label{eq:bloch_density_matrix}
\end{equation}
where
\begin{equation}
\mathbf{r}(t)
=
\left(
r_x(t),r_y(t),r_z(t)
\right)
\end{equation}
The Bloch-vector components are given by
\begin{equation}
r_q(t)=\langle\sigma_q\rangle_t
\qquad
q\in\{x,y,z\}
\label{eq:bloch_components}
\end{equation}
where $\id$ denotes the $2\times2$ identity matrix and $\sigma_q$ with
$q \in \{x,y,z\}$ are the Pauli matrices that define the $SU(2)$
algebra on the qubit Hilbert space.

The qubit is initialized in the north-pole localized state,
\begin{equation}
\rho(0)
=
|+\ z\rangle\langle+ \ z|,
\qquad
\mathbf{r}(0)
=
(0,0,1)
\label{eq:initial_state}
\end{equation}

Following the kinematic mixed-state GP of Tong \textit{et al.},
\cite{tong2004kinematic} let the instantaneous spectral decomposition
of the reduced density matrix be
\begin{equation}
\rho(t)
=
\sum_k
\lambda_k(t)
|\varphi_k(t)\rangle
\langle\varphi_k(t)|
\end{equation}
The gauge-invariant kinematic geometric phase accumulated between
$0$ and $t$ is then
\begin{equation}
\begin{aligned}
\gamma_D(t)
=
\arg\Bigg[
\sum_{k}&
\sqrt{\lambda_k(0)\lambda_k(t)}
\,
\langle\varphi_k(0)|\varphi_k(t)\rangle
\\[-2pt]
&\times
\exp\left(
-\int_0^t
\langle\varphi_k(\tau)|
\dot{\varphi}_k(\tau)\rangle
\,\dd\tau
\right)
\Bigg]
\end{aligned}
\label{eq:tong_general}
\end{equation}
Because the initial state in Eq.~\eqref{eq:initial_state} is pure,
only the eigenstate initially carrying unit population contributes to
Eq.~\eqref{eq:tong_general}. For the corresponding instantaneous
eigenvector written in the north-patch gauge as
\begin{equation}
|\varphi_{+z}(t)\rangle
=
\begin{pmatrix}
\cos[\theta(t)/2]\\
e^{i\phi(t)}\sin[\theta(t)/2]
\end{pmatrix}
\label{eq:north_eigenvector}
\end{equation}
one obtains
\begin{equation}
\langle\varphi_{+z}(t)|
\dot{\varphi}_{+z}(t)\rangle
=
\frac{i}{2}
\left[
1-\cos\theta(t)
\right]
\dot{\phi}(t)
\label{eq:berry_connection_north}
\end{equation}
The endpoint overlap
$\langle\varphi_{+z}(0)|\varphi_{+z}(t)\rangle=\cos[\theta(t)/2]$ is real
and non-negative wherever the north-patch representation remains
regular. The Tong expression therefore reduces, modulo $2\pi$, to
\begin{equation}
\gamma_D(t)
=
-\frac{1}{2}
\int_0^t
\left[
1-\cos\theta(\tau)
\right]
\dot{\phi}(\tau)
\,\dd\tau
\label{eq:GP_driven_theta_phi}
\end{equation}
where
\begin{equation}
\cos\theta(t)
=
\frac{r_z(t)}{|\mathbf{r}(t)|},
\qquad
\phi(t)
=
\operatorname{atan2}
\left[
r_y(t),r_x(t)
\right]
\label{eq:theta_phi}
\end{equation}

The principal-branch representation of $\operatorname{atan2}$ lies in
$(-\pi,\pi]$ and therefore contains an artificial $2\pi$ branch cut. \cite{strichartz2003guide} 
Moreover, the north-patch coordinate representation in
Eq.~\eqref{eq:GP_driven_theta_phi} becomes singular at the south pole.
Accordingly, the geometric phases reported below are evaluated from
the gauge-invariant Tong construction in
Eq.~\eqref{eq:tong_general}, while
Eq.~\eqref{eq:GP_driven_theta_phi} provides the corresponding
Bloch-sphere representation wherever the north-patch coordinates are
regular. Equation~\eqref{eq:GP_driven_theta_phi} makes explicit that
the GP depends on both the polar excursion of the Bloch vector (given
by $1-\cos\theta(t)$) and its oriented azimuthal winding rate (given by
$\dot{\phi}(t)$). 

To quantify non-additive drive-bath effects, one can use the exclusion-inclusion principle to define,
\begin{equation}
\delta\gamma_D(t)
=
\gamma_D^{A,\Omega}(t)
-
\gamma_D^{A=0}(t)
-
\gamma_U^{A,\Omega}(t)
+
\gamma_0^U(t)
\label{eq:delta_GP}
\end{equation}
Because each GP is defined modulo $2\pi$, direct
subtraction of independently represented phases can contain arbitrary
integer multiples of $2\pi$ and therefore the non-additive
phase through its circular difference becomes useful,
\begin{equation}
\delta\gamma_D^{\mathrm{circ}}(t)
=
\operatorname{Arg}
\left\{
e^{
i\left(
\gamma_D^{A,\Omega}(t)
-
\gamma_D^{A=0}(t)
-
\gamma_U^{A,\Omega}(t)
+
\gamma_0^U(t)
\right)}
\right\}
\label{eq:delta_GP_circular}
\end{equation}
with $\delta\gamma_D^{\mathrm{circ}}\in(-\pi,\pi]$. Equivalently, the
branch-independent non-additive phase factor is
\begin{equation}
Z_\delta(t)
=
e^{
i\left[
\gamma_D^{A,\Omega}(t)
-
\gamma_D^{A=0}(t)
-
\gamma_U^{A,\Omega}(t)
+
\gamma_0^U(t)
\right]}
\label{eq:Z_delta}
\end{equation}

$Z_\delta(t)$ is invariant under
$\delta\gamma_D\rightarrow\delta\gamma_D+2\pi n, \ n \in \mathbb{Z} $. A value $Z_\delta=1$ indicates additivity of the drive-induced and bath-induced GP modifications with respect to the common bare unitary
reference ($\gamma_0^U$), whereas deviations from unity identify a cooperative drive-bath effects. It is useful to derive the unitary GP directly from the Bloch-vector
dynamics. One can show (see Appendix~\ref{app:unitary_gp_derivation}) that the unitary driven GP can be written as
\begin{equation}
\begin{aligned}
\gamma_U^{A,\Omega}(t)
={}&
-\frac{\Delta}{2}
\int_0^t
\frac{r_x(\tau)r_z(\tau)}
{1+r_z(\tau)}
\,\dd\tau
\\
&+
\frac{A}{2}
\int_0^t
\left[1-r_z(\tau)\right]
\cos(\Omega\tau)
\,\dd\tau 
\end{aligned}
\label{eq:unitary_GP_exact}
\end{equation}
Equation~\eqref{eq:unitary_GP_exact} is an exact kinematic expression
for the unitary driven GP in terms of the Bloch-vector dynamics. Short time expansion of Equation~\eqref{eq:unitary_GP_exact} detailed in Appendix~\ref{app:unitary_gp_derivation}
gives an analytic check for the numerical results presented in Section~\ref{sec:results}


\section{Floquet Picture of Drive--Bath Spectral Steering}
\label{sec:floquet_picture}


\subsection{Floquet dressing of the coherent dynamics}
\label{sec:floquet_dressing}

Because $H_S(t+T_D)=H_S(t)$, the coherent driven dynamics admits a
Floquet representation.\cite{sato2025floquet,grifoni1999dissipation,
magazzu2018asymptotic,mori2023floquet}.The corresponding Floquet states of $H_S$ can be written as
\begin{equation}
|\psi_\alpha(t)\rangle
=
e^{-i\varepsilon_\alpha t}
|u_\alpha(t)\rangle,
\quad
|u_\alpha(t+T_D)\rangle
=
|u_\alpha(t)\rangle
\label{eq:floquet_state}
\end{equation}
where $\ket{u_\alpha(t)}$ are time-periodic Floquet modes and
$\varepsilon_\alpha$ are the corresponding quasienergies satisfying 
the eigenvalue equation,
\begin{equation}
\left[H_S(t)-i\partial_t\right]\ket{u_\alpha(t)}
=
\varepsilon_\alpha\ket{u_\alpha(t)}
\end{equation}
To expose the harmonic structure generated by the longitudinal drive, the drive term can be removed explicitly through the time-dependent unitary transformation,
\begin{equation}
U_D(t)=e^{i\frac{A}{2\Omega}\sin(\Omega t)\sigma_z}
\label{eq:drive_transformation}
\end{equation}

The transformed system Hamiltonian becomes
\begin{equation}
\begin{aligned}
\widetilde{H}_S(t)
&=
U_D^\dagger(t)H_S(t)U_D(t)
-iU_D^\dagger(t)\dot{U}_D(t)
\\
&=
-\frac{\Delta}{2}
\left[
\cos\!\left(\kappa\sin(\Omega t)\right)\sigma_x
+
\sin\!\left(\kappa\sin(\Omega t)\right)\sigma_y
\right]
\end{aligned}
\label{eq:transformed_system}
\end{equation}
where $\kappa=A/\Omega$ is the dimensionless drive-dressing parameter. It directly controls the harmonic content of the drive reormalized Hamiltonian.

Using the Jacobi-Anger expansion, Eq.~\eqref{eq:transformed_system} can be expressed as a Fourier series in a Bessel function \cite{abramowitz1964handbook} basis as,
\begin{align}
\widetilde{H}_S(t) = -\frac{\Delta}{2}\Bigg\{
&
\left[
\mathcal{J}_0(\kappa)
+
2\sum_{m=1}^{\infty}
\mathcal{J}_{2m}(\kappa)
\cos(2m\Omega t)
\right]\sigma_x
\nonumber\\
&
+
2\sum_{m=0}^{\infty}
\mathcal{J}_{2m+1}(\kappa)
\sin[(2m+1)\Omega t]\sigma_y
\Bigg\}
\label{eq:bessel_decomposition}
\end{align}

The drive therefore restructures the bare tunneling dynamics into a
periodically rotating transverse field containing a hierarchy of
Floquet harmonics whose weights are controlled by the dressing
parameter $\kappa=A/\Omega$.

In the high-frequency limit ($\Omega/\Delta\gg 1$), the leading Floquet effective Hamiltonian is obtained by averaging over one drive period,
\begin{equation}
\begin{aligned}
H_F^{(0)}
&=
\frac{1}{T_D}
\int_0^{T_D}
\widetilde{H}_S(t)\,dt
=
-\frac{\Delta}{2}
\mathcal{J}_0(\kappa)\sigma_x
\\
\Delta_{\mathrm{eff}}
&=
\Delta\mathcal{J}_0(\kappa)
\end{aligned}
\label{eq:Delta_eff}
\end{equation}

In general the harmonic content of the drive renormalized system from Eq.~\eqref{eq:bessel_decomposition} can be classified into three distinct operational regimes
\begin{equation}
\begin{aligned}
\kappa &\ll 1
&&\text{: weak dressing}\\
\kappa &\sim 1
&&\text{: intermediate dressing}\\
\kappa &\gg 1
&&\text{: strong dressing}
\end{aligned}
\label{eq:dressing_regimes}
\end{equation}

For weak dressing, $\kappa\ll1$, the Bessel function can be expanded
as $\mathcal{J}_0(\kappa)=1-\frac{\kappa^2}{4}+\frac{\kappa^4}{64}+\mathcal{O}(\kappa^6)$. Thus, to leading nontrivial order,
\begin{equation}
\Delta_{\mathrm{eff}}
\simeq
\Delta
\left(
1-\frac{\kappa^2}{4}
\right)
\label{eq:weak_drive_Delta}
\end{equation}
In this regime the effective transverse field undergoes only a small
angular excursion, and the drive produces a perturbative reduction
of the bare tunneling amplitude.
\\
\\
For intermediate dressing, $\kappa\sim1$, the expansion in powers of
$\kappa$ is no longer sufficient and the full Bessel-function
dependence becomes important. In this regime, higher Fourier components in
Eq.~\eqref{eq:bessel_decomposition} can acquire appreciable weight,
so that multiple Floquet harmonics contribute to the dynamics.
\\
\\
For strong dressing, $\kappa\gg1$, the asymptotic form of the Bessel
function is $\mathcal{J}_0(\kappa)\simeq\sqrt{\frac{2}{\pi\kappa}}\cos\left(\kappa-\frac{\pi}{4}\right)$
Consequently,
\begin{equation}
\Delta_{\mathrm{eff}}
\simeq
\Delta
\sqrt{\frac{2}{\pi\kappa}}
\cos\left(
\kappa-\frac{\pi}{4}
\right)
\label{eq:strong_kappa_Delta}
\end{equation}

Strong dressing therefore produces both a suppression of the
effective tunneling and repeated sign changes of $\Delta_{\mathrm{eff}}$ as a function of $\kappa$. In such regimes rotating wave approximation (RWA) can then become inadequate, requiring a more complete Floquet treatment. \cite{coleman2024spectral} The numerical results reported here don't invoke RWA so they are valid across a broad range of $\kappa$ values. In high frequency drive limit, the zeros of $\mathcal{J}_0(\kappa)$ correspond to points at which the leading-order effective tunneling vanishes. The first zero occurs at $\kappa_1\simeq2.4048$. At these points, $\Delta_{\mathrm{eff}}=0$ corresponding to CDT \cite{grossmann1991coherent} due to dynamical stabilization of a localized initial state.\cite{makri1997stabilization} Previous work by Dong and Makri \cite{dong2004quantum,dong2004optimizing} provides useful mechanistic insight into intra-doublet and inter-doublet dynamics of a symmetric dissipative qubit under strong one-color driving by seeking quantized representations of the classical drive field to interpret drive induced renormalization of the bare system Hamiltonian. These ideas provide a useful framework for interpreting the drive amplitude dependence on the dissipative GP.


\subsection{Floquet spectral sampling of the environment}
\label{sec:spectral_steering}

Floquet-dressed transition frequencies have the general form
\begin{equation}
\omega_{\alpha\beta}^{(n)}
=
\varepsilon_\alpha
-
\varepsilon_\beta
+
n\Omega
\label{eq:floquet_transition_frequency}
\end{equation}
where $\varepsilon_\alpha$ and $\varepsilon_\beta$ are Floquet
quasienergies.

The environment can therefore induce transitions between
drive-dressed states through multiple Floquet sidebands. Such
drive-dependent dissipative dynamics has been studied in driven
spin-boson type systems. \cite{grifoni1999dissipation,magazzu2018asymptotic,cangemi2019dissipative}

Since the system-bath interaction given by
Eq.~\eqref{eq:system_bath_interaction} is proportional to $\sigma_z$,
it remains unchanged under the transformation defined by
Eq.~\eqref{eq:drive_transformation}, since
\begin{equation}
U_D^\dagger(t)\sigma_z U_D(t)
=
\sigma_z
\end{equation}

Consequently,
\begin{equation}
\widetilde{H}_{SB}
=
-\frac{1}{2}\sigma_z B_z
\label{eq:transformed_HSB}
\end{equation}

The drive therefore modifies the system dynamics while leaving the
physical system-bath coupling operator unchanged. However, by dressing the system it does change the bath induced transition rates, which in the weak system-bath coupling Floquet--Markov description
\cite{mickiewicz2026benchmarking} is given by a Fermi-Golden Rule (FGR) type expression,
\begin{equation}
\Gamma_{\alpha\rightarrow\beta}
\sim
\sum_n
\left|
\sigma_{z_{\alpha\beta}}^{(n)}
\right|^2
J
\left(
\left|
\varepsilon_\alpha
-
\varepsilon_\beta
+
n\Omega
\right|
\right)
\label{eq:floquet_rates}
\end{equation}
where $\sigma_{z_{\alpha\beta}}^{(n)}$ denotes the $n$th Fourier component of the Floquet matrix element of $\sigma_z$. Equation~\eqref{eq:floquet_rates} provides a
weak-coupling spectral interpretation of the numerically converged PT-TEMPO results by identifying the bath frequencies sampled by the Floquet-dressed system. One might anticipate GP protection against environmental deformation to occur when Floquet dressing redistributes dominant transition weight into weakly weighted spectral regions of the bath.

\begin{figure*}[t]
\centering

\includegraphics[width=0.95\textwidth]{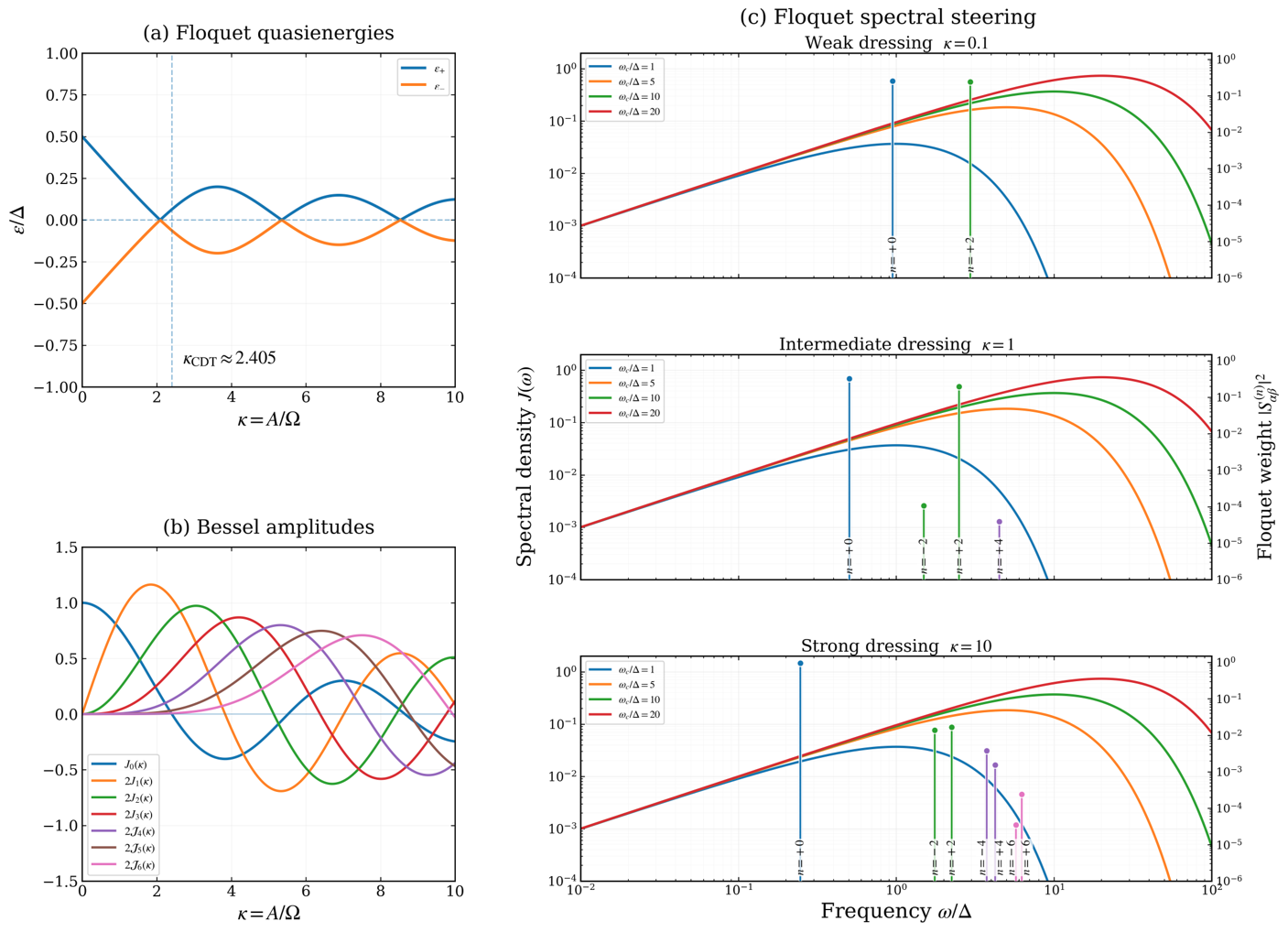}

\caption{\justifying
The mechanism underlying drive--bath spectral steering of the dissipative GP can be understood by looking at the quantities discussed below. Here the qubit is driven resonantly ($\Omega/\Delta=1$) and the drive dressing parameter is varied by simply changing the amplitude of the drive.
(a) Floquet quasienergy structure of the periodically driven symmetric
qubit as a function of the drive dressing parameter.
(b) Representative Fourier/Bessel weights characterizing the evolution
from weak to strong Floquet dressing.
(c) Floquet-dressed transition frequencies superimposed on
representative Ohmic bath spectral densities.
This figure illustrates the two key ingredients influencing the dissipative
GP: the drive-dependent Floquet transition weights and the intrinsic bath
spectral density available at the corresponding transition frequencies of the harmonically restructured qubit.
}
\label{fig:floquet_mechanism}

\end{figure*}

\begin{figure}[t]
\centering

\includegraphics[width=\columnwidth]{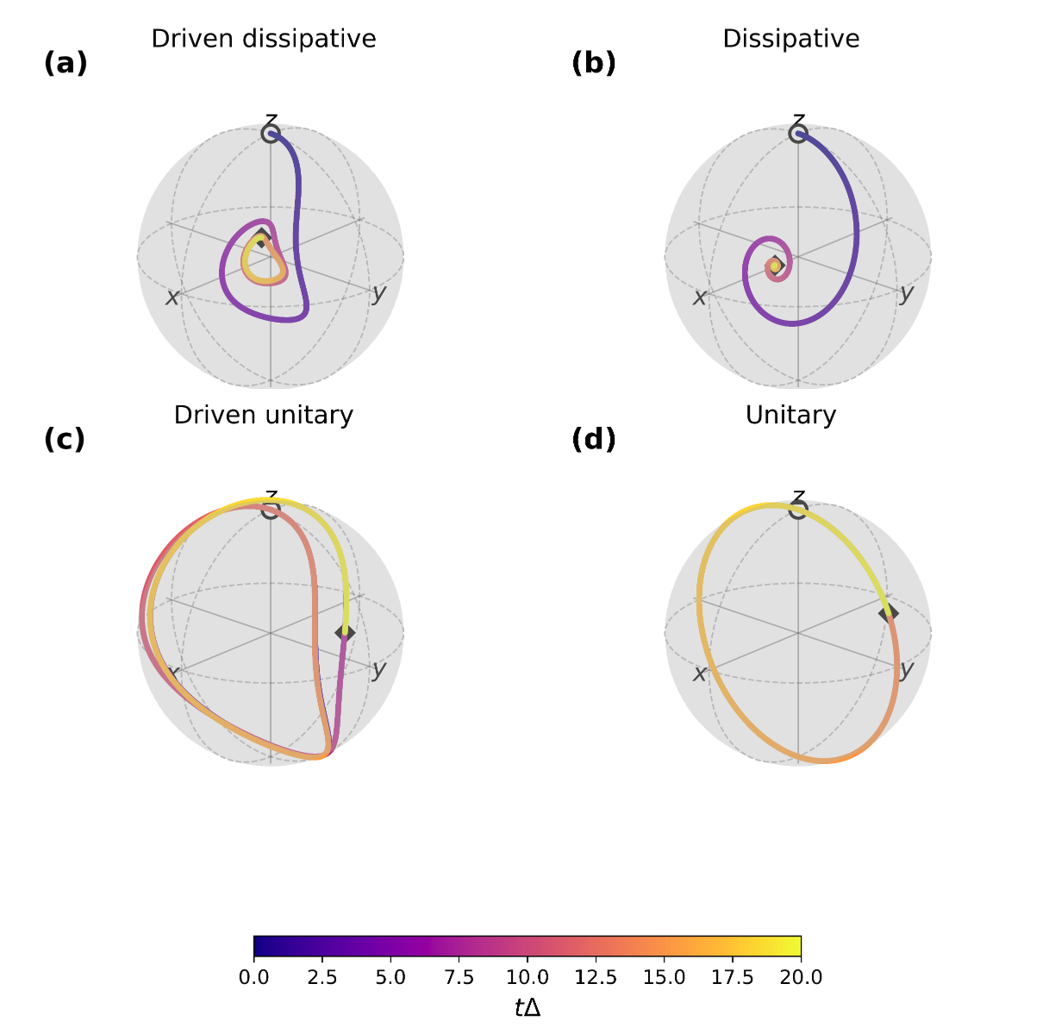}

\caption{\justifying
Bloch-sphere trajectories of the qubit for the
(a) driven dissipative, (b) undriven dissipative,
(c) driven unitary, and (d) undriven unitary dynamics.
The trajectories illustrate how periodic driving and environmental
dissipation deform the path followed by the Bloch vector on
the Bloch sphere relative to the undriven unitary reference.
For the dissipative dynamics in panels (a) and (b) the bath parameters
are $\omega_c/\Delta=20$, $\alpha=0.05$, $T/\Delta=1.5$, and
$t_{\mathrm{mem}}=7\tau_c$. For the driven dynamics in panels (a) and
(c) the drive parameters are $A/\Delta=1$ and $\Omega/\Delta=1$. Each trajectory is colored according to the dimensionless time $t\Delta$, from $t\Delta=0$ to $20$. The open circle and black diamond indicate the initial and final states, respectively. Dissipative trajectories are shown using the physical, unnormalized Bloch vector, such that motion into the interior of the sphere represents environmentally induced loss of purity.
}
\label{fig:bloch_trajectories}

\end{figure}

Thus, the drive can modify the environment induced deformation of GP through 
two related mechanisms. First, it modifies the coherent system dynamics through
Floquet dressing, whose strength and harmonic structure are controlled
by $\kappa=A/\Omega$. This produces a renormalization of the effective
tunneling amplitude, generates higher-order Floquet harmonics, and can
strongly suppress tunneling near the zeros of $\mathcal{J}_0(\kappa)$, as illustrated by Eq.~\eqref{eq:bessel_decomposition}. Second, the drive modifies the frequencies at which the dissipative environment couples to transitions between Floquet-dressed states. The resulting Floquet sidebands sample different regions of the bath spectral density. The competition between coherent Floquet dressing, higher order Floquet sidebands, and the frequency-dependent bath spectral weight therefore provides a natural framework for interpreting the drive-amplitude dependence of the
dissipative GP. These two effects are quantitatively illustrated in Fig.~\ref{fig:floquet_mechanism} and an illustrative set of Bloch trajectories is for all combinations of driving and dissipation are shown in Fig.~\ref{fig:bloch_trajectories}


\section{Results and Discussion}
\label{sec:results}

This section is organized as follows. First, the effect of drive
amplitude is examined under resonant conditions, spanning weak to
strong driving and highlighting the associated drive--bath
non-additivity. The corresponding Bloch trajectories illustrate
the geometric deformation induced by the environment. Next,
frequency and coupling-strength sweeps reveal how the one-cycle
geometric phase and its bath-induced shift depend on the driving
conditions and bath cutoff. Finally, a two-cycle echo protocol
is examined as a reversibility diagnostic, and its limited
connection to the Tong geometric phase is discussed.


\subsection{From weak to strong Floquet dressing under resonant driving}
\label{sec:amplitude_results}

At resonant driving, $\Omega=\Delta$, the dressing parameter
$\kappa=A/\Omega$ is numerically equal to the normalized drive
amplitude $A/\Delta$. The three amplitudes considered here,
$A/\Delta=0.1$, $1.0$, and $10.0$, therefore correspond directly
to weak, intermediate, and strong dressing regimes, respectively.

\begin{figure*}[t]
\centering

\includegraphics[width=0.95\textwidth]{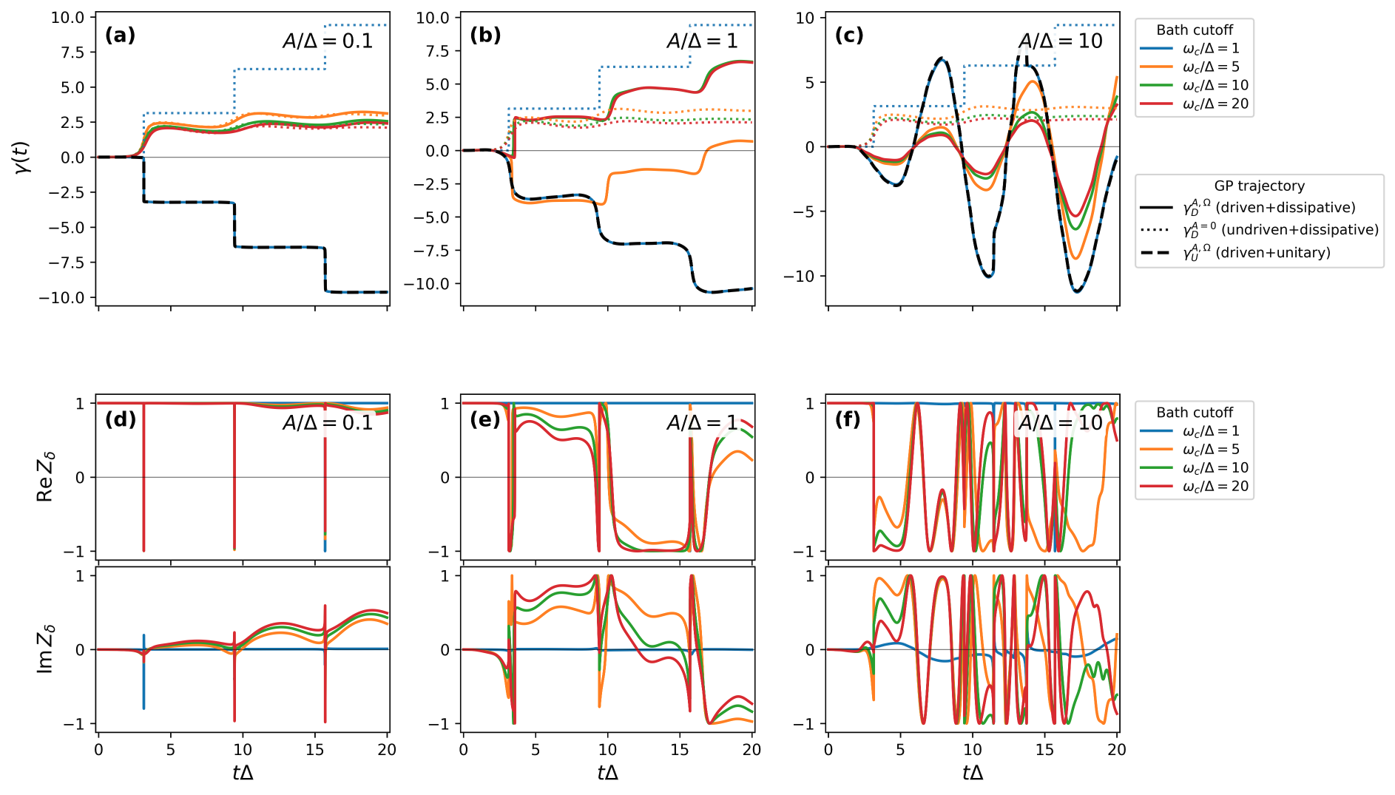}

\caption{\justifying
Crossover from weak to strong Floquet dressing under resonant driving.
Columns correspond to
(a,d) $A/\Delta=0.1$,
(b,e) $A/\Delta=1$, and
(c,f) $A/\Delta=10$,
with $\Omega/\Delta=1$.
The upper row compares the driven-dissipative geometric phase
$\gamma_D^{A,\Omega}$ with the driven-unitary reference
$\gamma_U^{A,\Omega}$ and the undriven-dissipative reference
$\gamma_D^{A=0}$ for bath cutoff frequencies
$\omega_c/\Delta\in\{1,5,10,20\}$.
The lower panels show the real and imaginary components of the
branch-independent non-additive phase factor $Z_\delta(t)$.
Weak driving leaves $Z_\delta$ close to unity, intermediate driving
produces pronounced bath cutoff sensitive deviations from unity, and strong driving generates rapid excursions with repeated phase winding.
The common bath parameters are
$\alpha=0.05$, $T/\Delta=1.5$, and
$t_{\mathrm{mem}}=7\tau_c$.
}
\label{fig:dressing_regimes}

\end{figure*}

Figure~\ref{fig:dressing_regimes} compares the geometric-phase (GP)
trajectories at resonant driving, $\Omega/\Delta=1$, for three amplitudes:
$A/\Delta=0.1$, $1$, and $10$. The upper panels show the driven
dissipative phase alongside the driven-unitary and undriven-dissipative
references for $\omega_c/\Delta \in \{1,5,10,20\}$. The lower panels
show the real and imaginary parts of the circular non-additivity measure
$Z_\delta$ defined in Eq.~\eqref{eq:Z_delta}. 

At the weakest amplitude [Figs.~\ref{fig:dressing_regimes}(a) and
\ref{fig:dressing_regimes}(d)], the dressing parameter is
$\kappa=A/\Omega=0.1$. The static component of the transformed
Hamiltonian remains close to its undriven value,
$\mathcal{J}_0(0.1)\simeq0.9975$ yet the driven
dissipative GP lies well away from that of the driven-unitary counterpart for 
most bath cutoffs with the exception of slow bath case with $\omega_c/\Delta = 1$.
This difference shows that weak driving does not imply a small
bath-induced GP shift. It does not, by itself, establish a large
non-additive effect as for much of the evolution window
$\operatorname{Re}Z_\delta$ remains close to unity and
$\operatorname{Im}Z_\delta$ remains comparatively small. 

At $A/\Delta=1$ [Figs.~\ref{fig:dressing_regimes}(b) and
\ref{fig:dressing_regimes}(e)], the first few harmonics of the
transformed Hamiltonian have appreciable amplitude. The leading
high-frequency effective-Hamiltonian term is
$\Delta\mathcal{J}_0(1)\simeq0.7652\Delta$ and it only characterizes the static
Fourier component rather than providing a complete effective
tunneling renormalization. The cutoff dependence of the driven
dissipative GP becomes particularly pronounced. For
$\omega_c/\Delta=5$, its trajectory is closer to the driven-unitary
reference than those for several other cutoffs over parts of the
evolution, despite the continued presence of bath coupling.
Meanwhile, the broad departures of both $\operatorname{Re}Z_\delta$
and $\operatorname{Im}Z_\delta$ from $(1,0)$ show a sustained
non-additive contribution. The ordering with $\omega_c$ is
non monotonic, i.e. increasing the cutoff does not produce a progressive
increase or decrease in the GP non-additivity.

Strong driving, $A/\Delta=10$ [Figs.~\ref{fig:dressing_regimes}(c)
and \ref{fig:dressing_regimes}(f)] leads to several Fourier components of
the transformed Hamiltonian be appreciable and produces substantially larger
oscillations of the GP. The corresponding $Z_\delta$ trajectories span
much of the unit circle, indicating large and rapidly varying
non-additive drive-bath effects.

The amplitude dependence for resonant driving suggests that even a weak
or intermediate amplitude radiation field can be utilized to control the GP. Figure ~\ref{fig:bloch_trajectories_resonant driving} shows the drive dissipative (a-d) and driven unitary (i) Bloch trajectories for resonant driving along with the undriven dissipative trajectories (e-h).
It is seen that cutoff frequency of the bath strongly impacts the bath induced deformation of the driven unitary trajectory. A particularly interesting case is for 
$\omega_c/\Delta = 1$ where driven dissipative (a) and driven unitary (i) are very close to one another. This is consequently also manifested in the driven dissipative GP tracking the driven unitary GP as seen in Figure ~\ref{fig:dressing_regimes}.

\subsection{Drive-frequency dependence}
\label{sec:frequency_results}

Having examined the crossover from weak to strong Floquet dressing under
resonant driving, the role of the drive frequency is examined by fixing the
drive amplitude at $A/\Delta=1$ and varying $\Omega/\Delta$. In this case,
the Floquet dressing parameter $\kappa=A/\Omega$ is controlled directly by
the drive frequency. At low $\Omega$, $\kappa$ is large and the system is
strongly Floquet dressed, whereas increasing $\Omega$ reduces $\kappa$ and
progressively suppresses the relative importance of higher-order harmonic
components. To compare different drive frequencies on an equal cycle-by-cycle basis,
the GP accumulated during exactly one period of the drive field is
considered. Defining $T_D=2\pi/\Omega$, the dissipative one-cycle GP is
\begin{equation}
    \gamma_D^{(1)}(\Omega,\omega_c)
    =
    \gamma_D^{A,\Omega}
    \left(
        T_D=\frac{2\pi}{\Omega},
        \omega_c
    \right)
    \label{eq:one_cycle_GP}
\end{equation}
with the corresponding driven-unitary quantity
\begin{equation}
    \gamma_U^{(1)}(\Omega)
    =
    \gamma_U^{A,\Omega}
    \left(
        T_D=\frac{2\pi}{\Omega}
    \right)
    \label{eq:one_cycle_GP_unitary}
\end{equation}
This single cycle GP avoids comparing trajectories after different
numbers of completed drive periods when $\Omega$ is varied, although the
physical duration of one cycle, $T_D=2\pi/\Omega$, remains intrinsically
frequency dependent. The frequency sweep is explored for the range
$0.4\leq\Omega/\Delta\leq10$ and thus it includes both above and below resonance conditions. 
The lower bound $\Omega/\Delta=0.4$ ensures
that at least one complete drive period is contained within the propagation
window $t\Delta\leq20$.

\begin{figure*}[t]
\centering

\includegraphics[width=0.94\textwidth]{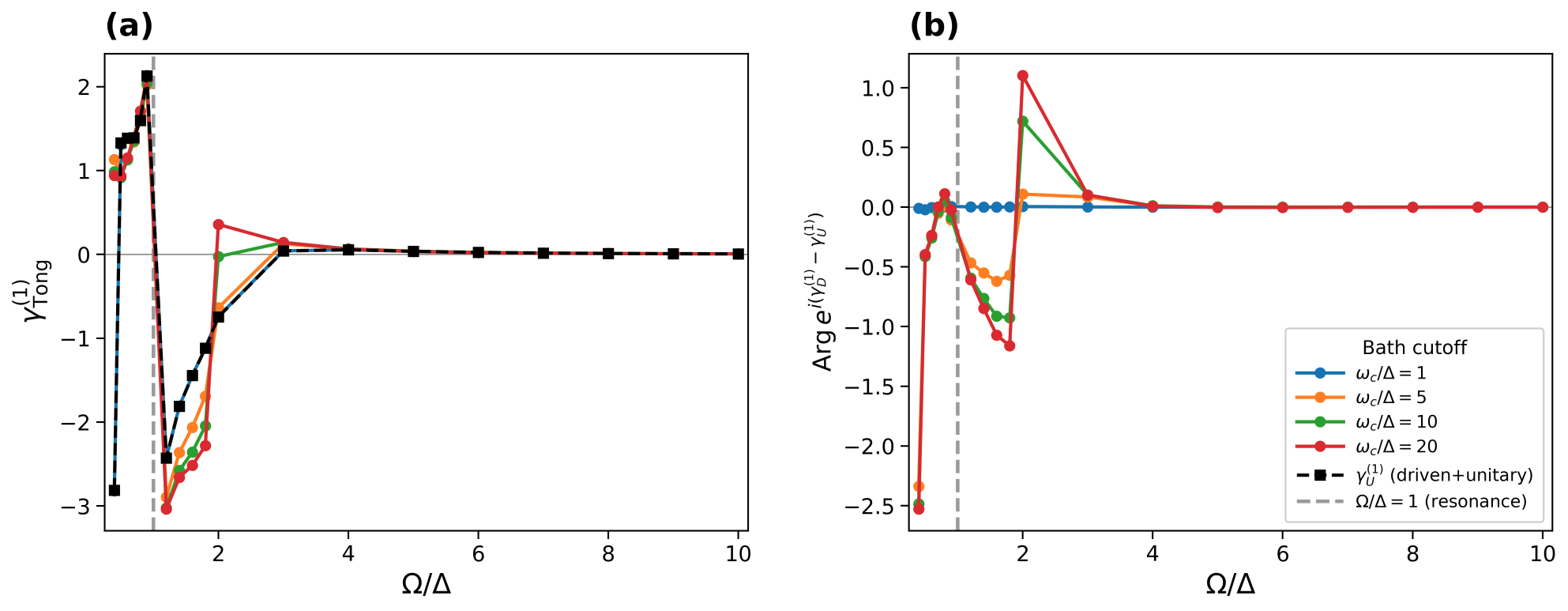}

\caption{\justifying
Drive-frequency dependence of the one-cycle geometric phase for fixed
driving amplitude $A/\Delta=1$ and bath cutoff frequencies
$\omega_c/\Delta=1,5,10,$ and $20$.
(a) One-cycle dissipative geometric phase
$\gamma_D^{(1)}(\Omega,\omega_c)$ as a function of normalized drive
frequency $\Omega/\Delta$. The corresponding driven-unitary result
$\gamma_U^{(1)}(\Omega)$ is shown for comparison.
(b) Circular phase shift due to bath-induced modification of the one-cycle geometric phase,
for the same bath cutoff frequencies.
The common bath parameters are $\alpha=0.05$, $T/\Delta=1.5$, and
$t_{\mathrm{mem}}=7\tau_c$. The vertical dashed line in both panels
indicates the resonance condition $\Omega/\Delta=1$.
}
\label{fig:frequency_sweep}

\end{figure*}

Figure~\ref{fig:frequency_sweep}(a) shows a pronounced nonmonotonic
frequency dependence of the one-cycle GP. In the sub-resonant regime,
$\gamma_D^{(1)}$ changes rapidly with increasing $\Omega$ and becomes
strongly enhanced on approaching the resonance. Above resonance, 
the one-cycle GP changes substantially by $\Omega/\Delta=2$ and subsequently approaches zero as the drive frequency is increased further. Importantly, a similar broad
non monotonic frequency structure is already present in the driven-unitary
reference. The overall drive frequency dependence is therefore primarily an intrinsic
feature of the coherently driven dynamics, while the environment modifies
this underlying response by deforming the coherent trajectory.

\begin{figure*}[t]
\centering

\includegraphics[width=0.94\textwidth]{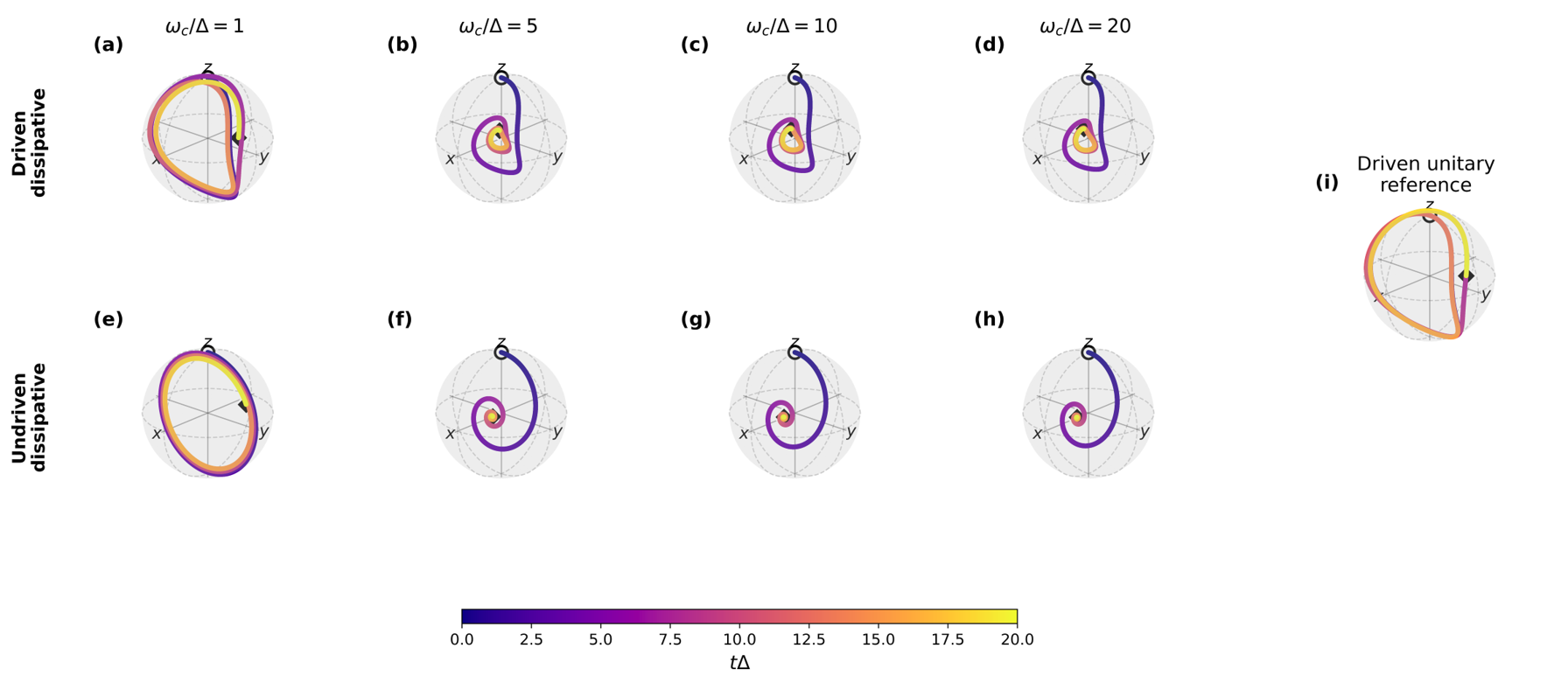}

\caption{\justifying
Bloch-sphere trajectories for fixed driving amplitude $A/\Delta=1$, drive frequency $\Omega/\Delta=1$, and bath cutoff frequencies $\omega_c/\Delta=1$, $5$, $10$, and $20$. (a)--(d) Driven dissipative trajectories for the respective bath cutoff frequencies. (e)--(h) Corresponding undriven dissipative trajectories. (i) Driven-unitary trajectory, which is independent of $\omega_c$, shown as a common reference. The common bath parameters are $\alpha=0.05$, $T/\Delta=1.5$, and $t_{\mathrm{mem}}=7\tau_c$. Each trajectory is colored according to the dimensionless time $t\Delta$, from $t\Delta=0$ to $20$. The open circle and black diamond indicate the initial and final states, respectively. Dissipative trajectories are shown using the physical, unnormalized Bloch vector, such that motion into the interior of the sphere represents environmentally induced loss of purity.
}
\label{fig:bloch_trajectories_resonant driving}

\end{figure*}

The bath-induced modification of the one-cycle GP is quantified by
the circular phase difference
\begin{equation}
\begin{aligned}
\Delta\gamma_{\mathrm{bath}}^{(1)}
(\Omega,\omega_c)
=
\operatorname{Arg}\Bigl\{
e^{
i\bigl(
\gamma_D^{(1)}(\Omega,\omega_c)
-
\gamma_U^{(1)}(\Omega)
\bigr)
}
\Bigr\}
\end{aligned}
\label{eq:one_cycle_bath_difference}
\end{equation}
As shown in Fig.~\ref{fig:frequency_sweep}(b), this deformation depends
strongly on both the drive frequency \emph{and} the bath cutoff. For some
combinations of $\Omega$ and $\omega_c$, the driven dissipative GP
$\gamma_D^{(1)}(\Omega,\omega_c)$, remains close to its unitary counterpart,
$\gamma_U^{(1)}(\Omega)$ whereas other combinations produce a substantially larger bath-induced
deformation. The bath induced deformation is not simply monotonic in $\omega_c$,
indicating that changing the bath cutoff does not produce a uniform
renormalization of the coherent dynamics as also evident from (e-h) in Fig.~\ref{fig:bloch_trajectories_resonant driving}.
It is worth noting that low frequency driving regimes doesn't necessarily define a proper
adiabatic limit for an OQS.\cite{viotti2023geometric} Such limits require further consideration \cite{venuti2016adiabaticity} 
since a complex interplay of three competing timescales: system ($\tau_S \sim \Delta^{-1}$), bath ($\tau_B \sim \omega_c^{-1}$)
and drive ($\tau_D \sim \Omega^{-1}$) must be considered for adiabaticity.

The particularly large bath-induced differences found at the lowest drive
frequencies should, however, be interpreted together with the increasing
cycle duration. Conversely, in the high-frequency regime the one-cycle interval becomes
progressively shorter, and both the dissipative and unitary one-cycle GPs approach zero.
The high-frequency suppression of $\gamma^{(1)}$ therefore characterizes
the geometric phase generated per drive cycle and should not, by itself,
be interpreted as the disappearance of environmental effects over a fixed
physical observation time.

Taken together, the frequency sweep separates two complementary effects.
The dominant nonmonotonic dependence of the one-cycle GP on $\Omega$ is
already encoded in the coherent driven dynamics and it changes as the Floquet
dressing parameter $\kappa=A/\Omega$ is varied. Adding dissipation to this then 
distorts this intrinsic underlying response, with the magnitude and sign of the
bath-induced correction depending on both $\Omega$ and $\omega_c$. Drive
frequency therefore provides a means of controlling both the coherent
geometric response and the manner in which the environment modifies that
response.


\subsection{Dependence on system--bath coupling strength}
\label{sec:alpha_results}

\begin{figure*}[t]
\centering

\includegraphics[width=0.94\textwidth]{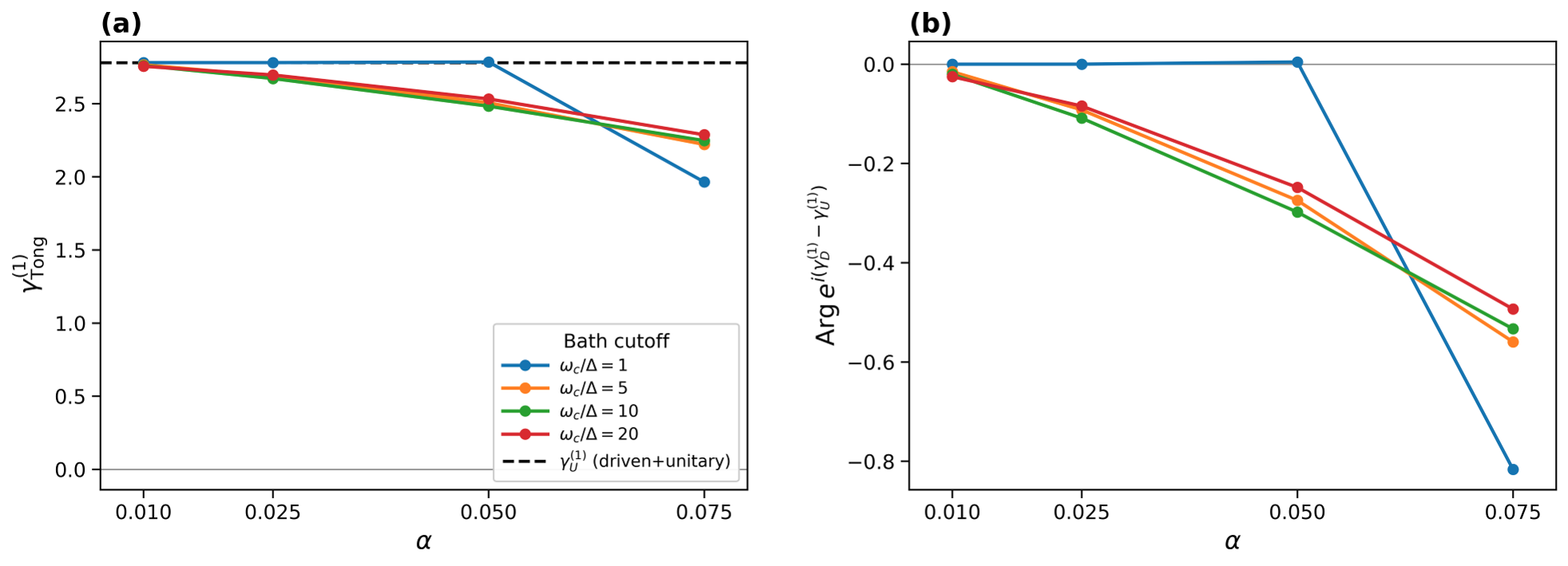}

\caption{\justifying
One-cycle geometric phase as a function of system--bath coupling
strength for $\omega_c/\Delta=1$, $5$, $10$, and $20$.
The left panel shows the driven dissipative phase
$\gamma_{D,\mathrm{Tong}}^{(1)}$ the black dashed line is the
driven-unitary result $\gamma_{U,\mathrm{Tong}}^{(1)}$.
The right panel shows the corresponding bath-induced phase shift,
$\Delta\gamma_{\mathrm{bath}}^{(1)}
=\arg\!\left[
e^{i(\gamma_{D,\mathrm{Tong}}^{(1)}
-\gamma_{U,\mathrm{Tong}}^{(1)})}
\right]$.
The common parameters across all curves are $A/\Delta=\Omega/\Delta=1$ and $T/\Delta=1.5$.
}

\label{fig:alpha_sweep}

\end{figure*}

The overall magnitude of the environmental deformation of the coherent trajectory can be varied independently through the Ohmic coupling strength $\alpha$ (at fixed $\omega_c$). Under such conditions, increasing $\alpha$ enhances the bath spectral weight while leaving the drive-generated Floquet frequencies determined  primarily by $A$ and $\Omega$. The $\alpha$ dependence therefore provides a useful insight into drive-bath competition. To distinguish this overall coupling-strength dependence from the frequency structure of the bath, Fig.~\ref{fig:alpha_sweep} compares the one cycle GP and its bath induced deformation under resonant driving for the same $\alpha$ sweep for representative cutoff frequencies.


In the weak-coupling regime, $\alpha=0.010$, the driven dissipative GP lies
close to the driven-unitary reference for all bath cutoffs giving corresponding values
of $\Delta\gamma_{\mathrm{bath}}^{(1)} \sim 0$. This shows that the one-cycle trajectory remains predominantly controlled by the resonant drive at weak coupling. As $\alpha$ is increased, the  driven dissipative GP for $\omega_c/\Delta=5$, $10$, and $20$ move progressively below the unitary value and consequently the bath-induced phase shifts become increasingly negative. Thus, for this set of cutoffs, strengthening the system--bath interaction produces a smooth and approximately monotonic deformation of the driven geometric phase with $\alpha$ for a given bath cutoff.

The behavior at $\omega_c/\Delta=1$ is qualitatively different. For this
slow-bath case, the dissipative phase remains nearly indistinguishable from
the unitary result up to $\alpha=0.050$, after which it decreases sharply at
$\alpha=0.075$. This is also reflected in the corresponding value of $\Delta\gamma_{\mathrm{bath}}^{(1)}$.
This pronounced nonlinear response should not be interpreted as a discontinuous transition, rather it indicates that
the one-cycle geometric phase becomes strongly sensitive to the dissipative
trajectory once the coupling exceeds the weak-coupling regime. Because
$\omega_c/\Delta=1$ also corresponds to the longest bath correlation time
considered here, this point provides a particularly stringent case for the
numerical-convergence analysis. (see Appendix~\ref{app:convergence})

The cutoff dependence is therefore not described by a simple ordering in
which increasing $\omega_c$ always increases the magnitude of the phase
deformation. At intermediate coupling, the largest shifts among the
higher-cutoff baths occur near $\omega_c/\Delta=10$, whereas at
$\alpha=0.075$ the strongest response is instead obtained for
$\omega_c/\Delta=1$. This nonmonotonic ordering reflects the fact that the
geometric phase depends on the complete reduced-state trajectory rather than
only on the integrated strength of the spectral density. Changing
$\omega_c$ redistributes the bath spectral weight relative to the
drive-generated Floquet transition frequencies and simultaneously changes
the bath memory time. Increasing $\alpha$ then amplifies the effect of this
cutoff-dependent spectral sampling.


\subsection{Echo-based interferometric protocol}
\label{sec:interferometric_protocol}

A two-cycle spin-echo protocol inspired by geometric-phase (GP)
interferometry in monitored open quantum systems is considered
\cite{Viotti2023}. Here, the protocol provides an operational diagnostic
of environmentally modified reversibility. Its persistence probability
is compared with a separately calculated Tong phase to assess the
limitations of identifying the echo signal with the dissipative GP obtained from RDM.

For the symmetric DSBM considered throughout this work, the system
Hamiltonian in Eq.~\eqref{eq:driven_system_hamiltonian} satisfies
\begin{equation}
H_S(0)=H_S(T_D)
=-\frac{\Delta}{2}\sigma_x-\frac{A}{2}\sigma_z
\label{eq:end_point_driven_system_hamiltonian}
\end{equation}
The effective field at the beginning and end of each drive cycle
therefore lies in the $x$--$z$ plane. The state
\begin{equation}
\ket{\psi_{\mathrm{I}}}
=
\ket{+\ y}
=
\frac{1}{\sqrt{2}}
\left(
\ket{+\ z}+i\ket{-\ z}
\right)
\label{eq:interferometer_initial_state}
\end{equation}
has equal-magnitude projections onto the two instantaneous eigenstates
of $H_S(0)$ and is used as the balanced interferometric input state.

During the first stage, the system is propagated from $t=0$ to
$t=T_D$ under Eq.~\eqref{eq:driven_system_hamiltonian}. At $t=T_D$,
an \emph{instantaneous} $\pi$ rotation about the $y$ axis is applied,
\begin{equation}
U_{\pi_y}
=
\exp\left(-i\frac{\pi}{2}\sigma_y\right)
=
-i\sigma_y
\label{eq:pi_y_pulse}
\end{equation}
Under this rotation,
\begin{equation}
\begin{aligned}
U_{\pi_y}\sigma_xU_{\pi_y}^{\dagger}&=-\sigma_x\\
U_{\pi_y}\sigma_zU_{\pi_y}^{\dagger}&=-\sigma_z
\end{aligned}
\label{eq:pi_y_operator_transformation}
\end{equation}
and consequently
\begin{equation}
U_{\pi_y}H_S(t)U_{\pi_y}^{\dagger}
=
-H_S(t)
\label{eq:pi_y_hamiltonian_swap}
\end{equation}
at every time. In particular, the pulse interchanges the two
instantaneous eigenstates of the endpoint Hamiltonian. After the
pulse, the system is propagated for a second period under the
temporally reversed control schedule,
\begin{equation}
H_{\mathrm{rev}}(s)
=
H_S(T_D-s),
\qquad
0\leq s\leq T_D
\label{eq:reversed_hamiltonian}
\end{equation}

An exact benchmark follows in the absence of the bath. Let
$U_{\mathrm{fwd}}$ and $U_{\mathrm{rev}}$ denote the isolated-system
propagators for the first and second stages, respectively. Temporal
reversal of the schedule, together with
Eq.~\eqref{eq:pi_y_hamiltonian_swap}, gives
\begin{equation}
U_{\mathrm{rev}}
=
U_{\pi_y}U_{\mathrm{fwd}}^{\dagger}U_{\pi_y}^{\dagger}
\end{equation}
and hence
\begin{equation}
U_{\mathrm{rev}}U_{\pi_y}U_{\mathrm{fwd}}
=
U_{\pi_y}
\label{eq:exact_unitary_echo}
\end{equation}
Since $U_{\pi_y}\ket{+y}=-i\ket{+y}$, the complete isolated
sequence returns the chosen input state up to an overall phase.
Its persistence probability is therefore exactly unity. This result
does not require adiabatic evolution and does not depend on the
one-cycle Tong phase.

The bath remains coupled throughout $0\leq t\leq2T_D$. In the
PT-TEMPO calculation, the $\pi_y$ pulse is inserted as an instantaneous
control superoperator,
\begin{equation}
\mathcal{U}_{\pi_y}[\rho]
=
U_{\pi_y}\rho U_{\pi_y}^{\dagger}
\label{eq:pulse_superoperator}
\end{equation}
within a single continuous process-tensor contraction. The process
tensor is not restarted at $t=T_D$, so system--bath correlations
and memory across the pulse is fully retained (within the memory cutoff limits).

The isolated-system identity in
Eq.~\eqref{eq:exact_unitary_echo} does not extend directly to the
combined system and bath. Writing
$H(t)=H_S(t)+H_{SB}+H_B$, the system pulse gives
\begin{equation}
U_{\pi_y}H(t)U_{\pi_y}^{\dagger}
=
-H_S(t)-H_{SB}+H_B
\label{eq:echo_total_hamiltonian_transformation}
\end{equation}
where the pulse acts trivially on the bath. The bath Hamiltonian
is not reversed, and the second stage therefore does not undo
the complete system--bath evolution.

At the end of the protocol, the persistence probability is
\begin{equation}
\begin{aligned}
P_{\mathrm{echo}}
&=
\bra{\psi_{\mathrm{I}}}
\rho_{\mathrm{echo}}(2T_D)
\ket{\psi_{\mathrm{I}}}\\
&=
\operatorname{Tr}
\left[
\rho_{\mathrm{I}}\rho_{\mathrm{echo}}(2T_D)
\right]
\end{aligned}
\label{eq:echo_persistence}
\end{equation}
where $\rho_{\mathrm{I}}=\ket{\psi_{\mathrm{I}}}
\bra{\psi_{\mathrm{I}}}$. For the chosen input, this is equivalently
\begin{equation}
P_{\mathrm{echo}}
=
\frac{1+r_y^{\mathrm{echo}}(2T_D)}{2}
\label{eq:echo_bloch_projection}
\end{equation}
The quantity thus measures recovery of the initial-state Bloch
projection and is sensitive to both coherent evolution and its
environmental induced deformation.

For comparison with the angle parametrization used in GP
interferometry \cite{Viotti2023}, define
\begin{equation}
P_{\mathrm{echo}}
=
\cos^{2}\left(2\phi_{\mathrm{echo}}^{(\mathrm{p})}\right)
\label{eq:echo_angle_definition}
\end{equation}
with
\begin{equation}
\begin{aligned}
\phi_{\mathrm{echo}}^{(\mathrm{p})}
&=
\frac{1}{2}
\cos^{-1}\left(\sqrt{P_{\mathrm{echo}}}\right)\\
0\leq\phi_{\mathrm{echo}}^{(\mathrm{p})}
&\leq\frac{\pi}{4}
\end{aligned}
\label{eq:principal_echo_angle}
\end{equation}
This angle is a parametrization of persistence rather than an
independently measured phase. Environmental changes in populations
and coherence can change its value without representing a geometric
phase shift. Furthermore, the mapping
$P=\cos^{2}(2\phi)$ cannot resolve the sign or the $\pi/2$ branch
of a putative phase.

For comparison, the upper instantaneous eigenstate
$\ket{\psi_{+}(0)}$ of
Eq.~\eqref{eq:end_point_driven_system_hamiltonian} is propagated
independently for one drive period. The GP of the resulting
reduced-state trajectory is calculated using the gauge-invariant
kinematic expression of Tong \textit{et al.}
\cite{tong2004kinematic} in Eq.~\eqref{eq:tong_general}.
This reference calculation contains no intermediate $\pi$ pulse.

A comparison probability motivated by the ideal geometric-echo
relation is defined as
\begin{equation}
P_{\mathrm{GP}}
=
\cos^{2}
\left(
2\gamma_{\mathrm{Tong}}^{(+)}
\right)
\label{eq:ideal_gp_probability}
\end{equation}
For the present protocol, this is a Tong-based reference rather
than an established prediction for the measured persistence.
Their difference is
\begin{equation}
\delta P_{\mathrm{int}}
=
P_{\mathrm{echo}}
-
\cos^{2}
\left(
2\gamma_{\mathrm{Tong}}^{(+)}
\right)
\label{eq:interferometric_mismatch}
\end{equation}
A small $\delta P_{\mathrm{int}}$ indicates agreement between the
two probabilities but does not establish faithful GP encoding,
because the phase-to-probability mapping is many-to-one and
agreement can occur without a direct measurement relation.
Conversely, a finite mismatch does not necessarily happen due to
non-Markovian effects, it can already occur in the isolated system 
and can also arise from Markovian relaxation or pure dephasing.

\begin{figure*}[t]
\centering

\includegraphics[width=0.48\textwidth]{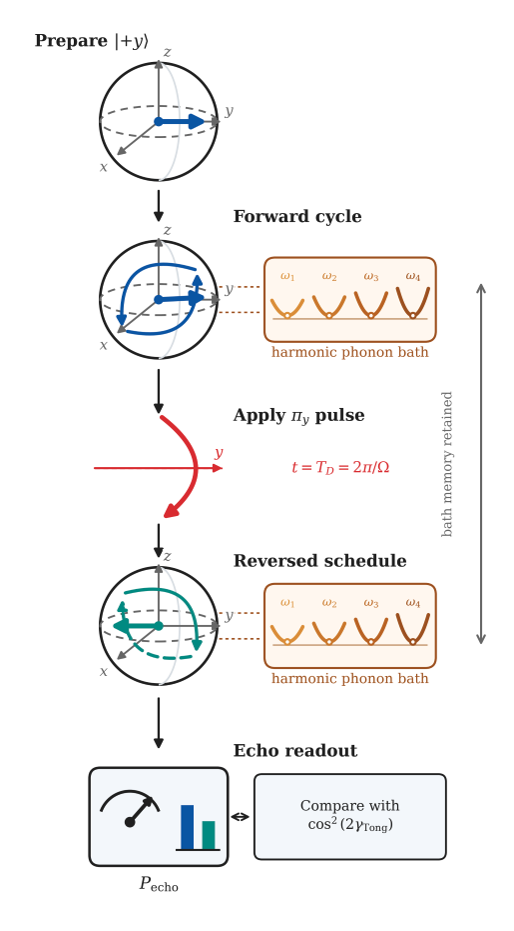}

\caption{\justifying
Visual depiction of the two-cycle interferometric protocol used to assess
the operational accessibility of the open-system geometric phase. The qubit
is prepared in the balanced superposition
$\ket{\psi_{\mathrm{I}}}=\ket{+ \ y}$ and propagated for one drive period
$T_D=2\pi/\Omega$ while continuously coupled to the harmonic phonon bath.
At $t=T_D$, an instantaneous $\pi_y$ pulse interchanges the two instantaneous
eigenstate branches. The qubit is then propagated for a second drive period
under the nominally reversed Hamiltonian schedule. A single process tensor is
retained over the complete interval $0\leq t\leq2T_D$, preserving bath memory
and system--bath correlations across the pulse. The final echo persistence
probability $P_{\mathrm{echo}}$ is measured relative to the initial state and
compared with $\cos^{2}(2\gamma_{\mathrm{Tong}})$, where
$\gamma_{\mathrm{Tong}}$ is obtained from a separate one-cycle evolution
without the intermediate $\pi_y$ pulse. Because the longitudinal cosine drive
traces a self-retracing path in Hamiltonian-parameter space, this comparison
is used as an operational diagnostic rather than as an assumed direct
measurement of the kinematic geometric phase.
}
\label{fig:interferometric_protocol_schematic}

\end{figure*}

For the longitudinal cosine drive in
Eq.~\eqref{eq:drive_hamiltonian},
\begin{equation}
\cos\left[\Omega(T_D-s)\right]
=
\cos(\Omega s)
\end{equation}
so that
\begin{equation}
H_{\mathrm{rev}}(s)
=
H_S(T_D-s)
=
H_S(s)
\label{eq:self_retracing_drive}
\end{equation}
The control therefore exhibits a self-retracing path in
Hamiltonian-parameter space; reversing the schedule does not
produce a distinct oppositely oriented parameter loop. This
property, together with the exact isolated-system recovery in
Eq.~\eqref{eq:exact_unitary_echo}, prevents a general identification
of the persistence angle with the independently calculated
one-cycle Tong phase. Accordingly, $P_{\mathrm{echo}}$,
$P_{\mathrm{GP}}$, and $\delta P_{\mathrm{int}}$ are reported
separately. Their comparison distinguishes environmentally
modified echo recovery from the kinematic GP of the
reference ($\pi$ pulse free) reduced-state trajectory. 
The two-cycle protocol is shown schematically in
Fig.~\ref{fig:interferometric_protocol_schematic}.


\begin{figure*}[t]
\centering

\includegraphics[width=0.97\textwidth]{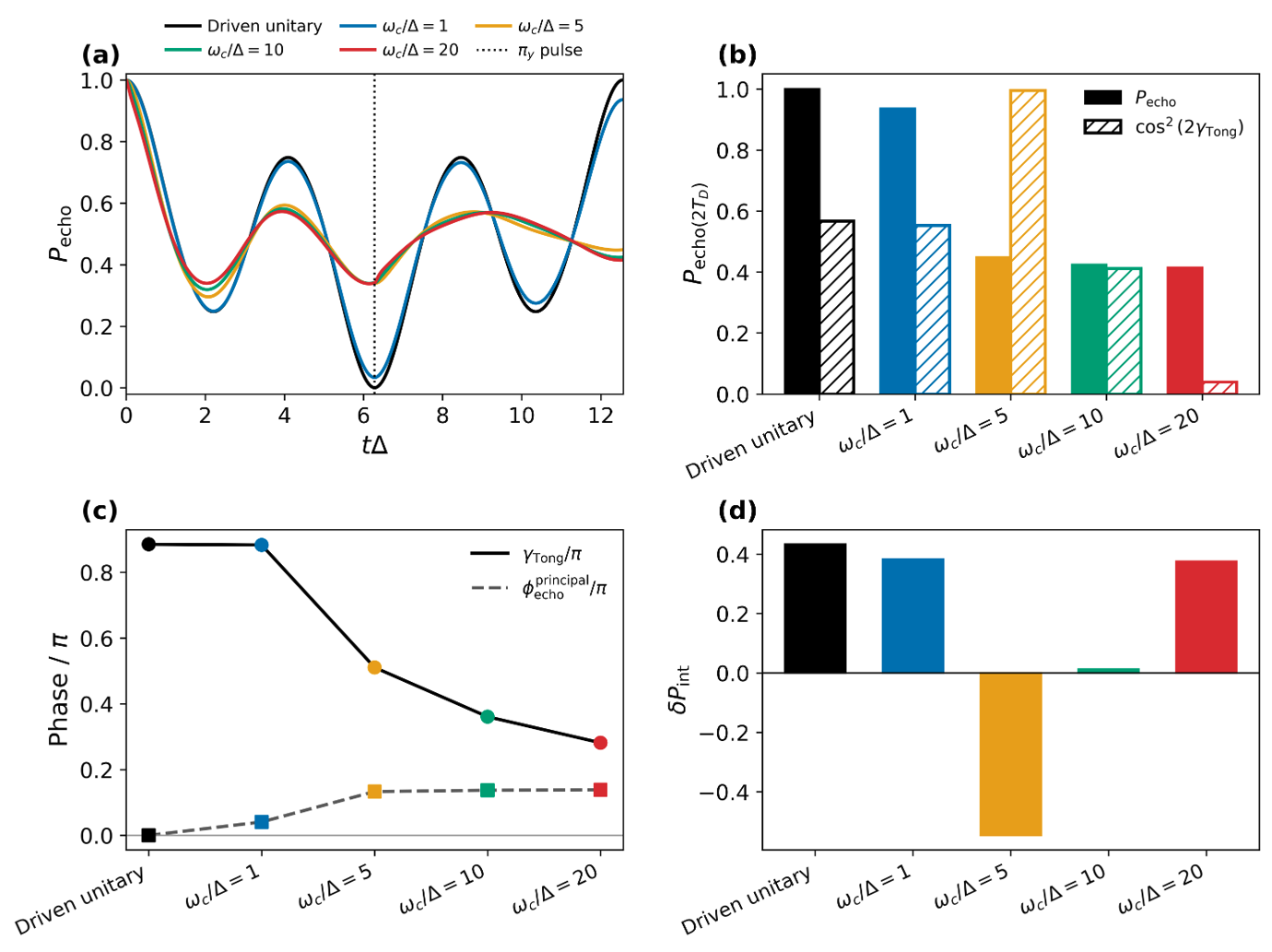}

\caption{\justifying
Two-cycle echo diagnostic for the longitudinally driven dissipative qubit.
The initial state $\ket{\psi_{\mathrm{I}}}=\ket{+ \ y}$ evolves for one
period $T_D=2\pi/\Omega$, followed by an instantaneous $\pi_y$ pulse
and a second period under the reversed schedule, which coincides with
the original cosine drive.
(a) Persistence probability
$P(t)=\operatorname{Tr}[\rho_{\mathrm{I}}\rho(t)]$
for unitary and dissipative dynamics with
$\omega_c/\Delta=1,5,10,$ and $20$; the dotted line marks the pulse.
(b) Final persistence $P_{\mathrm{echo}}$ and the Tong-based comparison
$P_{\mathrm{GP}}=\cos^2(2\gamma_{\mathrm{Tong}})$.
The Tong phase is calculated independently over one cycle from the
upper instantaneous eigenstate of $H_S(0)$, without a pulse.
(c) $\gamma_{\mathrm{Tong}}/\pi$ and the persistence-derived angle
$\phi_{\mathrm{echo}}^{(\mathrm{p})}/\pi$, where
$\phi_{\mathrm{echo}}^{(\mathrm{p})}
=\tfrac12\arccos\sqrt{P_{\mathrm{echo}}}$.
(d) Difference $\delta P_{\mathrm{int}}=P_{\mathrm{echo}}-P_{\mathrm{GP}}$.
The isolated protocol gives exact recovery, whereas bath coupling
suppresses recovery with a nonmonotonic cutoff dependence.
Neither the persistence-derived angle nor its agreement with the
Tong-based comparison establishes a direct GP measurement.
Parameters: $A/\Delta=\Omega/\Delta=1$, $\alpha=0.05$,
and $T/\Delta=1.5$.
}
\label{fig:interferometric_gp_results}

\end{figure*}

Fig.~\ref{fig:interferometric_gp_results} compares the two-cycle echo
protocol with the independently calculated one-cycle Tong geometric phase.
The driven-unitary evolution exhibits nearly perfect echo recovery,
$P_{\mathrm{echo}}\simeq1$, whereas
$\cos^{2}(2\gamma_{\mathrm{Tong}})$ remains substantially smaller. The
discrepancy therefore persists in the absence of environmental effects
and originates primarily from the self-retracing nature of the longitudinal
cosine protocol. Coupling to the bath suppresses the echo recovery and
produces a strongly nonmonotonic dependence on the cutoff frequency. In
particular, the near equality of the two probabilities at
$\omega_c/\Delta=10$ does not correspond to equality between the Tong phase
and the principal echo angle, reflecting the sign and branch information
lost under the mapping $P=\cos^{2}(2\phi)$. These results demonstrate that
the bath-modified geometric phase and interferometric reversibility constitute
distinct diagnostics of the driven open-system evolution.

\section{Summary and Outlook}
\label{sec:summary}

The geometric phase (GP) of a periodically driven symmetric
spin-boson model was investigated using PT-TEMPO reduced-state
dynamics and the gauge-invariant kinematic formulation of Tong
\textit{et al.} Comparisons with driven-unitary and
undriven-dissipative references, together with a circular
non-additivity measure, characterize how periodic driving and
dissipation jointly modify the GP.

Under resonant driving, increasing the amplitude produces a
crossover from weak modulation to substantial changes in the
phase trajectory and pronounced drive--bath non-additivity.
The bath-cutoff dependence is generally non monotonic, so increasing
the cutoff does not uniformly increase or decrease environmental induced
phase deformation. 

The one-cycle GP exhibits pronounced non monotonic frequency
dependence, with a similar broad structure already present in the
driven-unitary reference. The environment modifies this coherent
response in a cutoff-dependent manner. At weak system--bath
coupling, the dissipative phase approaches its unitary counterpart.
Increasing the coupling generally produces larger deviations,
although their sensitivity and ordering depend on the bath cutoff.
Periodic driving therefore reduces bath-induced GP deformation
in some parameter regimes while enhancing it in others.

These trends are interpreted qualitatively in terms of Floquet
dressing and changes in the spectral overlap between the driven
system and its environment. The Floquet-transition picture is
motivated by weak-coupling Floquet theory and provides a framework
for interpreting the simulations; it is not the dynamical
approximation underlying the PT-TEMPO calculations. PT-TEMPO
includes environmental memory without imposing weak-coupling or
Markov approximations.

The two-cycle echo provides a complementary reversibility
diagnostic. For the chosen input state, the isolated-system
propagator identity gives exact echo recovery, irrespective of
the independently calculated one-cycle Tong phase. Bath coupling
modifies this recovery, but agreement or disagreement with a
Tong-based probability comparison does not establish GP encoding
or identify non-Markovian effects.

Several limitations delimit the present conclusions. The
calculations employ an Ohmic environment with an exponential
cutoff and longitudinal system--bath coupling through the same
coordinate as the external field. Structured vibrational
environments, sub- or super-Ohmic spectral densities may produce
qualitatively different GP dynamics. The principal
GP calculations also consider monochromatic semiclassical
driving and an initially localized pure state, the echo uses
a separate superposition input. Although regimes of reduced
bath-induced phase deformation are identified, the study does
not establish a universal optimal-control protocol or a direct
experimental measurement scheme for the mixed-state GP.

Several extensions follow naturally. Optimal-control methods
could generate shaped fields that minimize environmental GP
deformation or approach a prescribed phase within a fixed time
window. Quantizing the applied field would introduce photon
statistics and field--system correlations as additional control
variables, allowing the effects of squeezing, antibunching, and
photon-number fluctuations to be examined. Broader classes of
initial states could clarify the state dependence of the results
and motivate geometric characterizations of the dynamical map
itself. Extensions to multilevel systems and more realistic dissipative
environments would further test the scope of the observed
drive--bath non-additivity.

Taken together, the results demonstrate that driving conditions
and bath spectral structure jointly determine the dissipative
GP response. The phase provides a trajectory-sensitive
characterization of this interplay, while Floquet dressing offers
a qualitative interpretation of its parameter dependence.
These findings motivate further exploration of field-based
control of environmental geometric-phase deformation.

\section*{Funding}
This research received no specific grant from any funding agency in the public, commercial, or not-for-profit sectors.
\section*{Author Contributions}
Chirag Arora: Conceptualization; Methodology; Validation; Formal analysis; Investigation; Visualization; Writing – original draft; Writing – review and editing.
\section*{Declaration of Generative Artificial Intelligence Use}

GPT-6 Sol Light was used to assist with literature searches, proofreading, and refinement of the manuscript's graphical presentation. The author independently evaluated and verified all scientific content, calculations, results, and conclusions.
\section*{Acknowledgments}
I would like to thank Prof. David F. Coker and Prof. Amartya Bose for their support and encouragement, along with careful reading of the manuscript and many edifying discussions.


\appendix


\section{Driven unitary geometric phase}
\label{app:unitary_gp_derivation}

It is useful to derive the unitary GP directly from the Bloch-vector
dynamics. Using the effective field in Eq.~\eqref{eq:effective_field}
the bath free driven Hamiltonian Eq.~\eqref{eq:driven_system_hamiltonian} can be written as
\begin{equation}
H_S(t)
=
-\frac{1}{2}\mathcal{B}(t)\cdot\boldsymbol{\sigma}
\label{eq:unitary_H_B}
\end{equation}
The von Neumann equation then gives
\begin{equation}
\dot{\mathbf r}
=
\mathbf r\times\mathcal{B}(t)
=
-\mathcal{B}(t)\times\mathbf r
\label{eq:unitary_bloch_correct}
\end{equation}
Thus, the Bloch vector precesses about the instantaneous effective
field $\mathbf{\mathcal{B}}(t)$ with the corresponding equations of motion 
\begin{align}
\dot{r}_x(t)
&=
A\cos(\Omega t)r_y(t)
\label{eq:unitary_bloch_x}
\\
\dot{r}_y(t)
&=
\Delta r_z(t)
-
A\cos(\Omega t)r_x(t)
\label{eq:unitary_bloch_y}
\\
\dot{r}_z(t)
&=
-\Delta r_y(t)
\label{eq:unitary_bloch_z}
\end{align}
These equations follow directly from the driven system Hamiltonian
in Eq.~\eqref{eq:driven_system_hamiltonian}.

The azimuthal velocity can be obtained directly from the Cartesian
Bloch-vector components,
\begin{equation}
\dot{\phi}(t)
=
\frac{r_x(t)\dot{r}_y(t)-r_y(t)\dot{r}_x(t)}
{r_x^2(t)+r_y^2(t)}
\label{eq:unitary_phidot_cartesian}
\end{equation}
Substitution of Eqs.~\eqref{eq:unitary_bloch_x} and
\eqref{eq:unitary_bloch_y} gives
\begin{equation}
\dot{\phi}(t)
=
\frac{
\Delta r_x(t)r_z(t)
-
A\cos(\Omega t)
\left[r_x^2(t)+r_y^2(t)\right]
}{
r_x^2(t)+r_y^2(t)
}
\end{equation}
Since the unitary evolution preserves $|\mathbf r|=1$,
\begin{equation}
r_x^2(t)+r_y^2(t)=1-r_z^2(t)
\end{equation}
and therefore
\begin{equation}
\dot{\phi}(t)
=
\Delta\frac{r_x(t)r_z(t)}
{1-r_z^2(t)}
-
A\cos(\Omega t)
\label{eq:unitary_phidot}
\end{equation}

For the unitary evolution, $\cos\theta= r_z$. Substitution of
Eq.~\eqref{eq:unitary_phidot} into the kinematic GP expression
Eq.~\eqref{eq:GP_driven_theta_phi} gives
\begin{equation}
\begin{aligned}
\gamma_U^{A,\Omega}(t)
=
-\frac{1}{2}
\int_0^t
\left[1-r_z(\tau)\right]
\left[
\frac{\Delta r_x(\tau)r_z(\tau)}
{1-r_z^2(\tau)}
-
A\cos(\Omega\tau)
\right]
\,\dd\tau
\end{aligned}
\end{equation}
Using
\begin{equation}
\frac{1-r_z}{1-r_z^2}
=
\frac{1}{1+r_z}
\end{equation}
this can be written as
\begin{equation}
\begin{aligned}
\gamma_U^{A,\Omega}(t)
={}&
-\frac{\Delta}{2}
\int_0^t
\frac{r_x(\tau)r_z(\tau)}
{1+r_z(\tau)}
\,\dd\tau
\\
&+
\frac{A}{2}
\int_0^t
\left[1-r_z(\tau)\right]
\cos(\Omega\tau)
\,\dd\tau 
\end{aligned}
\label{eq:unitary_GP_exact}
\end{equation}
Equation~\eqref{eq:unitary_GP_exact} is an exact kinematic expression
for the unitary driven GP in terms of the Bloch-vector dynamics. 
The behavior at short times provides a useful analytic check on the
sign convention used for the GP. Starting from
$\mathbf r(0)=(0,0,1)$, expansion of
Eqs.~\eqref{eq:unitary_bloch_x}--\eqref{eq:unitary_bloch_z} about
$t=0$ gives
\begin{align}
r_x(t)
&=
\frac{A\Delta}{2}t^2
+
\mathcal O(t^4)
\\
r_y(t)
&=
\Delta t
+
\mathcal O(t^3)
\\
r_z(t)
&=
1-\frac{\Delta^2}{2}t^2
+
\mathcal O(t^4)
\end{align}
Because the initial state $\rho(0)=|+ \ z\rangle\langle+ \ z|$ lies at the north pole of the Bloch sphere, the azimuthal angle $\phi=\operatorname{atan2}(r_y,r_x)$ is formally undefined at $t=0$. Its initial value is assigned from the continuous right-hand limit of the trajectory. For the present Hamiltonian, the short-time dynamics yield $r_y(t)=\Delta t+\mathcal{O}(t^3)$ and $r_x(t)=A\Delta t^2/2+\mathcal{O}(t^4)$, giving $\phi(0^+)=\pi/2$. Consequently, setting \(\phi(0)=\pi/2\) prior to phase unwrapping and numerical differentiation avoids the artificial discontinuity associated with the numerical evaluation of atan2(0,0). Nevertheless, its limiting behavior for $t>0$ follows from the Cartesian definition,
\begin{equation}
\phi(t)
=
\operatorname{atan2}[r_y(t),r_x(t)]
=
\frac{\pi}{2}
-\frac{A}{2}t
+
\mathcal O(t^3)
\end{equation}
so that
\begin{equation}
\dot{\phi}(t)
=
-\frac{A}{2}
+
\mathcal O(t^2)
\end{equation}
At the same time,
\begin{equation}
1-\cos\theta(t)
=
1-r_z(t)
=
\frac{\Delta^2}{2}t^2
+
\mathcal O(t^4)
\end{equation}
Consequently,
\begin{equation}
\dot{\gamma}_{U}^{A,\Omega}(t)
=
-\frac{1}{2}
\left[1-\cos\theta(t)\right]
\dot{\phi}(t)
=
\frac{A\Delta^2}{8}t^2
+
\mathcal O(t^4)
\end{equation}
and hence
\begin{equation}
\gamma_U^{A,\Omega}(t)
=
\frac{A\Delta^2}{24}t^3
+
\mathcal O(t^5)
\label{eq:unitary_GP_short_time}
\end{equation}


\section{PT-TEMPO convergence}
\label{app:convergence}

Numerical convergence of PT-TEMPO results was performed
with respect to the time step $dt$, bath-memory cutoff
$t_{\mathrm{mem}}$, and relative compression threshold
$\epsilon_{\mathrm{rel}}$ for resonant driving,
$A/\Delta=\Omega/\Delta=1$, of the most numerically demanding case
with $\alpha=0.075$, $\omega_c/\Delta=1$ and $T/\Delta=1.5$. 
Four numerical settings were compared:
\begin{enumerate}
\item Baseline:
$d t\Delta=0.01$,
$t_{\mathrm{mem}}=3/\omega_c$,
$\epsilon_{\mathrm{rel}}=10^{-4}$
\item Tolerance check:
$d t\Delta=0.01$,
$t_{\mathrm{mem}}=3/\omega_c$,
$\epsilon_{\mathrm{rel}}=10^{-5}$
\item Memory check:
$d t\Delta=0.01$,
$t_{\mathrm{mem}}=7/\omega_c$,
$\epsilon_{\mathrm{rel}}=10^{-5}$
\item Production:
$d t\Delta=0.0025$,
$t_{\mathrm{mem}}=7/\omega_c$,
$\epsilon_{\mathrm{rel}}=10^{-6}$
\end{enumerate}
The baseline and tolerance-check calculations isolate the
compression threshold, while the tolerance and memory checks
isolate the memory cutoff. The comparison between the memory
check and production jointly refines the time step and
compression threshold.

\begin{figure*}[t]
\centering
\includegraphics[width=0.95\textwidth]
{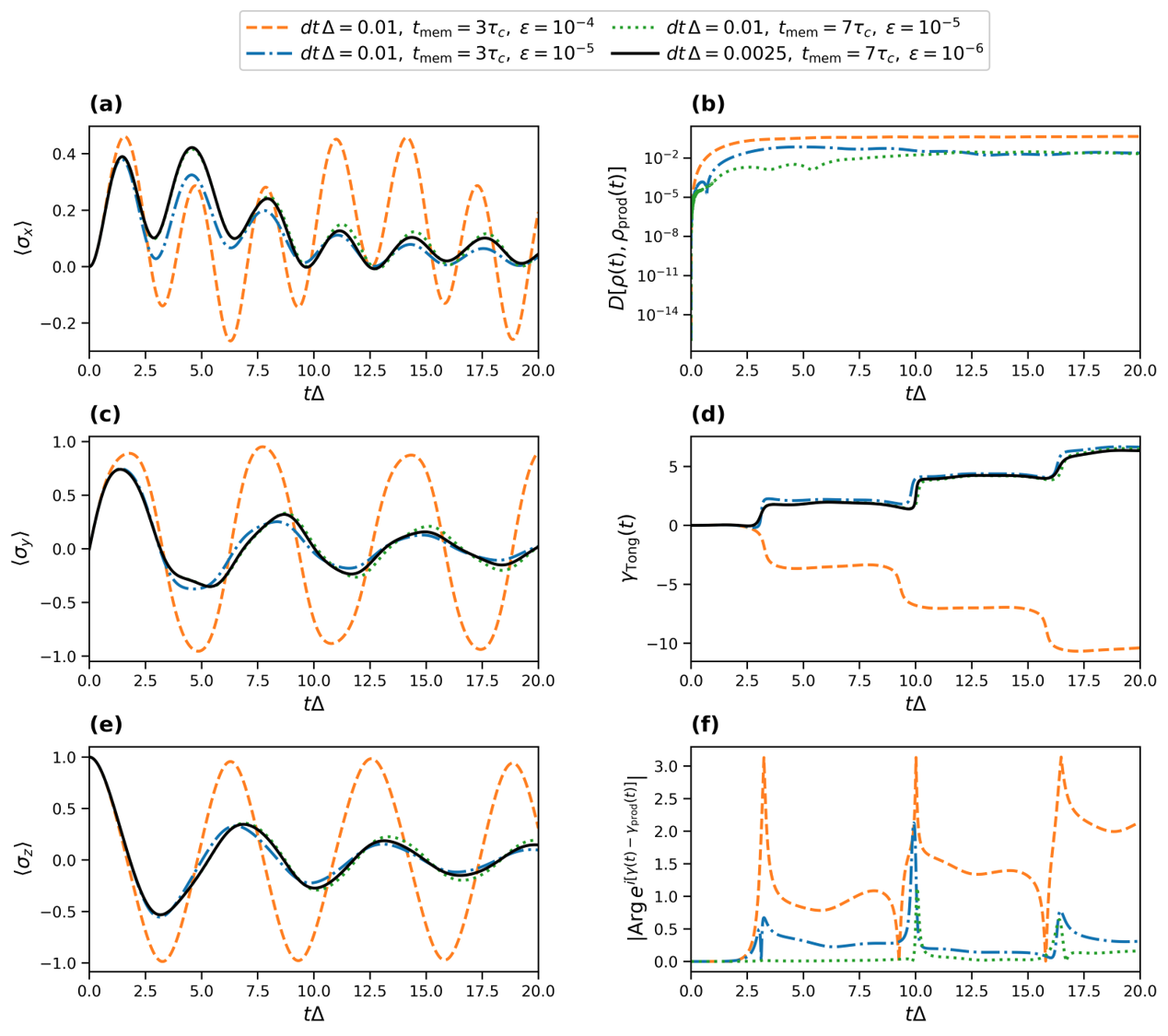}
\caption{PT-TEMPO numerical convergence assessment for the resonantly
driven spin-boson model at
$A/\Delta=\Omega/\Delta=1$, $\alpha=0.075$,
$\omega_c/\Delta=1$, and $T/\Delta=1.5$.
Panels (a), (c), and (e) show
$\langle\sigma_x\rangle$, $\langle\sigma_y\rangle$, and
$\langle\sigma_z\rangle$, respectively.
Panel (b) shows the trace distance from the production
density matrix on a logarithmic scale.
Panel (d) shows the continuously unwrapped Tong geometric
phase, and panel (f) shows its absolute circular difference
from production, defined in
Eq.~\eqref{eq:convergence_circular_error}.
Orange dashed, blue dash-dotted, green dotted, and black
solid curves denote the baseline, tolerance check, memory
check, and production settings, respectively.
}

\label{fig:pt_tempo_convergence}
\end{figure*}

Figure~\ref{fig:pt_tempo_convergence} compares the three Bloch
components and the gauge-invariant Tong geometric phase.
The production trajectory is used as the numerical reference,
rather than an exact solution. RDM discrepancies
are quantified by the trace distance,
\begin{equation}
D(t)=\frac{1}{2}
\left\|\rho_{\mathrm{test}}(t)
-\rho_{\mathrm{prod}}(t)\right\|_1
\label{eq:convergence_trace_distance}
\end{equation}
and geometric-phase discrepancies by the circular difference,
\begin{equation}
\epsilon_\gamma(t)=
\left|
\operatorname{Arg}
\exp\!\left\{
i[\gamma_{\mathrm{test}}(t)
-\gamma_{\mathrm{prod}}(t)]
\right\}
\right|
\label{eq:convergence_circular_error}
\end{equation}
As before, this definition removes differences arising solely from
integer multiples of $2\pi$.

The baseline setting produces substantial discrepancies in
both the Bloch dynamics and geometric phase. Tightening the
compression threshold markedly improves agreement, and
extending the memory cutoff further improves the Bloch
trajectories and phase agreement over much of the evolution.
Residual RDM differences remain, with the
memory-check trace distance reaching a few times $10^{-2}$.
The corresponding phase error is small between rapid phase
changes but exhibits localized peaks near $t\Delta\simeq10$ and $16.5$. 
These peaks coincide with slightly displaced rapid changes
in the calculated geometric phases. Since the Tong phase
is the argument of a complex amplitude, small trajectory
differences can produce enhanced phase sensitivity when
that amplitude approaches zero. 

\section*{References}

\bibliography{aipsamp}

\end{document}